\documentstyle[12pt]{article}

\def\today{\number\day\space
     \ifcase\month\or
       January\or February\or March\or April\or May\or June\or
       July\or August\or September\or October\or November\or December\fi
     \space\number\year}

\input epsf

\def\dal{\hbox{\hskip 0.5mm\hbox{\vrule width2.3mm height0.2mm
\vbox{\hrule width0.3mm height2.6mm}\hskip
-2.6mm \vbox{\hbox{\vrule width2.6mm height0.1mm}
\vskip -0.1mm\hrule width0.1mm height2.6mm}}\hskip 0.5mm}}
\newcommand{\be}{\begin{equation}}
\newcommand{\ee}{\end{equation}}
\newcommand{\bea}{\begin{eqnarray}}
\newcommand{\eea}{\end{eqnarray}}
\newcommand{\bdm}{\begin{displaymath}}
\newcommand{\edm}{\end{displaymath}}
\newcommand{\benu}{\begin{enumerate}}

\newcommand{\eenu}{\end{enumerate}}
\newcommand{\we}{\wedge}

\newcommand{\lap}{\Delta}

\newcommand{\daltil}{\tilde{\dal}}

\newcommand{\nabtil}{\tilde{\nabla}}

\newcommand{\gtil}{\tilde{g}}
\newcommand{\ghat}{\hat{g}}

\newcommand{\RR}{\mbox{$I \! \! R$}}
\newcommand{\gtens}{\mbox{\boldmath $g$}}

\newcommand{\ghattens}{\hat{\gtens}}
\newcommand{\gtiltens}{\tilde{\gtens}}
\newcommand{\Tmatrix}{\mbox{\boldmath $T$}}
\newcommand{\Mmatrix}{\mbox{\boldmath $M$}}
\newcommand{\Dmatrix}{\mbox{\boldmath $D$}}
\newcommand{\Pmatrix}{\mbox{\boldmath $P$}}
\newcommand{\Aop}{\mbox{\boldmath $A$}}

\newcommand{\dgtil}{\mbox{\boldmath $\delta$} \tilde{\mbox{\boldmath $g$}}}
\newcommand{\dw}{\mbox{\boldmath $\delta w$}}
\newcommand{\dx}{\mbox{\boldmath $\delta z$}}
\newcommand{\da}{\mbox{\boldmath $\delta a$}}
\newcommand{\db}{\mbox{\boldmath $\delta b$}}
\newcommand{\dc}{\mbox{\boldmath $\delta c$}}
\newcommand{\dmu}{\mbox{\boldmath $\delta \mu$}}
\newcommand{\dR}{\mbox{\boldmath $\delta R$}}
\newcommand{\dvec}{\mbox{\boldmath $\delta v$}}
\newcommand{\dwdot}{\mbox{\boldmath $\delta$} \dot{\mbox{\boldmath $w$}}}
\newcommand{\dbdot}{\mbox{\boldmath $\delta$} \dot{\mbox{\boldmath $b$}}}
\newcommand{\dmudot}{\mbox{\boldmath $\delta$} \dot{\mbox{\boldmath $\mu$}}}
\newcommand{\dRdot}{\mbox{\boldmath $\delta$} \dot{\mbox{\boldmath $R$}}}
\newcommand{\dzetabar}{\overline{\mbox{\boldmath $\zeta$}}}
\newcommand{\dzeta}{\mbox{\boldmath $\zeta$}}
\newcommand{\dddzeta}{\ddot{\mbox{\boldmath $\zeta$}}}
\newcommand{\dddw}{\mbox{\boldmath $\delta$} \ddot{\mbox{\boldmath $w$}}}
\newcommand{\dpi}{\mbox{\boldmath $\delta \varpi$}}
\newcommand{\dpidot}{\mbox{\boldmath $\delta$} \dot{\mbox{\boldmath $\varpi$}}}
\newcommand{\dAtil}{\mbox{\boldmath $\delta$} \tilde{\mbox{\boldmath $A$}}}
\newcommand{\dazer}{\mbox{\boldmath $\delta a_0$}}
\newcommand{\daone}{\mbox{\boldmath $\delta a_1$}}
\newcommand{\daoned}{\mbox{\boldmath $\delta$} \dot{\mbox{\boldmath $a_1$}}} 
\newcommand{\dxi}{\mbox{\boldmath $\xi$}}
\newcommand{\dchi}{\mbox{\boldmath $\chi$}}
\newcommand{\dphi}{\mbox{\boldmath $\phi$}}
\newcommand{\dpsi}{\mbox{\boldmath $\psi$}}
\newcommand{\delrho}{\Delta \rho}

\newcommand{\dfrac}{\frac{\dzeta}{\mu'}}

\newcommand{\Bpl}{{\mbox{\boldmath $B$}}^{+}_{\psi}}
\newcommand{\Bmi}{{\mbox{\boldmath $B$}}^{-}_{\psi}}
\newcommand{\Bpm}{{\mbox{\boldmath $B$}}^{\pm}_{\psi}}

\newcommand{\drhoone}{\frac{\partial}{\partial \rho}}
\newcommand{\drhotwo}{\frac{\partial^{2}}{\partial \rho ^{2}}}
\newcommand{\detapsi}{\mbox{\boldmath $\eta$}_{\psi}}
\newcommand{\e}{\mbox{e}}
\newcommand{\Lregn}{\Lambda_{\mbox{\scriptsize reg}}(n)}
\newcommand{\Lcritn}{\Lambda_{\mbox{\scriptsize crit}}(n)}

\begin{document}

\begin{titlepage}

\begin{flushright}
hep-th/9605166\\
ZU-TH 14/96\\
May 1996
\end{flushright}

\vfill

\begin{center}

{\LARGE{\bf{Stability Analysis of New}}}\\ 
{\LARGE{\bf{Solutions of the EYM system}}}\\
{\LARGE{\bf{with Cosmological Constant}}}

\vfill

{\bf{O. Brodbeck, M. Heusler, 
G. Lavrelashvili\footnote{On leave of absence from 
Tbilisi Mathematical Institute, 380093 Tbilisi, Georgia},}}\\
{\bf{N. Straumann and M.S. Volkov}}

\vfill

{\it{Institut f\"ur Theoretische Physik, Universit\"at Z\"urich}}\\
{\it{Winterthurerstr. 190, CH--8057 Z\"urich, Switzerland}}

\vfill

{\Large{\bf{Abstract}}}

\end{center}

\vfill

\noindent
We analyze the stability properties
of the purely magnetic, static solutions to the 
Einstein--Yang--Mills equations with cosmological constant.
It is shown that all three classes of solutions found in
a recent study are unstable under spherical
perturbations. Specifically, we argue that the
configurations have $n$ unstable modes in each parity
sector, where $n$ is the number of nodes of the magnetic 
Yang--Mills amplitude of the background solution.

The ``sphaleron--like'' instabilities (odd parity modes)
decouple from the gravitational perturbations. They are
obtained from a regular Schr\"odinger equation after
a supersymmetric transformation. 

The body of the work is devoted to the fluctuations with 
even parity. The main difficulty arises because the Schwarzschild 
gauge -- which is usually imposed to eliminate
the gravitational perturbations from the Yang--Mills 
equation -- is not regular for solutions with compact spatial
topology. In order to overcome this problem, we derive a gauge
invariant formalism by virtue of which the unphysical (gauge) modes
can be isolated. 

\vfill

\end{titlepage}

\section{Introduction}

A previous numerical analysis \cite{VSLHB} has revealed the following 
features of the purely magnetic solutions to the spherically symmetric
Einstein--Yang--Mills (EYM) equations with positive cosmological
constant $\Lambda$:
The static configurations fall into three classes, where the solutions
in each class are characterized by the values of $\Lambda$ and the
number of zeroes, $n$, of the magnetic YM amplitude $w$.
For sufficiently small values, $\Lambda < \Lcritn$, the solutions
asymptotically approach the deSitter geometry and can be viewed as
Bartnik--McKinnon (BK) solitons \cite{BK} surrounded by a cosmological 
horizon.
For a set of values $\Lregn$ exceeding $\Lcritn$, the configurations
have the topology $\RR \we S^{3}$, where the ground state ($n=1$)
is the Einstein Universe with constant YM energy density.
Between $\Lcritn$ and $\Lregn$ there exists a discrete family of
``bag of gold'' solutions, that is, of configurations with horizon
and ``finite size''.

This paper is devoted to the stability properties of the solutions
found in \cite{VSLHB} with respect to spherically symmetric 
fluctuations.
The metric $\gtens$ of a spherically symmetric spacetime is described
in terms of the metric $\gtiltens$ of a $2$--dimensional 
pseudo--Riemannian manifold $\tilde{M}$ and a function 
$R$ on $\tilde{M}$, such that
$\gtens = \gtiltens + R^{2} d \Omega^{2}$.

The stability analysis of solutions for which $R$ has no critical
points is considerably simplified by the fact that $R$ can be used
as a coordinate on $\tilde{M}$.
Moreover, it turns out to be sufficient to analyze the
perturbations in the Schwarzschild gauge, $\delta R = 0$.
The linearized Einstein equations then assume the form of
{\it constraint\/} equations, by virtue of which one can express the 
metric perturbations $\delta \gtiltens$ in terms of the fluctuations of 
the matter fields.
In this way, one always ends up with a set of fluctuation equations for 
the matter fields in standard form \cite{BHS}.
For the BK solitons \cite{BK} and the corresponding black hole
solutions \cite{KM}, \cite{VG-BH}, \cite{B}, the set of pulsation
equations therefore reduces to a single regular Schr\"odinger equation
for the YM amplitude $w$. The number of bound states of this
equation eventually determines the number of unstable modes
\cite{SZ}, \cite{ZS} (in the even parity sector).

In the present paper we are mainly interested in the stability
properties of solutions for which $R$ {\it does\/} have a local extremum.
In this case, the Schwarzschild gauge, $\delta R = 0$, is not everywhere 
regular and the above procedure can therefore not be readily adopted.
In fact, for $\delta R \neq 0$, not all of the linearized Einstein 
equations have the form of constraints. Hence, one obtains a coupled 
system of pulsation equations, including both matter {\it and\/} metric
perturbations. Although this system is self--adjoint, the differential
operator is {\it not hyperbolic\/}, reflecting the fact that the equations 
contain unphysical degrees of freedom (gauge modes).

Studying the gauge dependence reveals that the
fluctuation equations actually contain only one degree of freedom which 
is of physical nature. In fact, it turns out to be possible to derive a
pulsation equation for the gauge invariant quantity $\zeta$ associated
with the physical degree of freedom. The procedure generalizes the 
method described above for the Schwarzschild gauge in a gauge invariant
way.

Although the pulsation equation for $\zeta$ is a standard
Schr\"odinger equation, it has the flaw that the potential is unbounded.
In fact, it turns out that the equator
(i.e., the critical point of $R$) is a {\it regular singular\/} point 
of the Sturm--Liouville equation for $\zeta$. A closer look reveals 
that the so--called limit point case occurs. 
``Weyl's alternative'' (see, e.g. \cite{Book})
then implies that there exists exactly one 
essentially self--adjoint realization of the pulsation operator, and 
that $\zeta$ must vanish at the equator for this realization.
However, there exist analytic solutions which do not
vanish at the equator and, nevertheless, give rise to perfectly 
regular metric fluctuations.

In order to reconstruct the metric and matter perturbations from
the gauge invariant quantity $\zeta$, we consider a class of gauges for
which all fluctuations can be obtained by solving a set of 
{\it ordinary\/} differential equations. A particular gauge of
this type is the Schwarzschild gauge, which is, however, not regular at
the equator. A {\it globally regular\/} gauge is obtained by 
considering {\it conformal\/} perturbations of the spatial part of the 
metric. In this way one finally obtains a 
{\it regular set of metric and matter fluctuations from each 
regular solution of the gauge invariant pulsation equation}.

The entire procedure is nicely illustrated for the lowest ($n=1$) 
compact solution, since all steps can be performed 
analytically in this case. The gauge invariant equation has one bound 
state, corresponding to an unstable mode. A numerical analysis 
reveals that the higher compact solutions ($n>1$) have, as expected, 
exactly $n$ unstable (gravitational) modes. 

The stability analysis described above applies to the family of 
{\it compact\/} solutions, parametrized by $n$ and $\Lregn$. 
The symmetry properties of these configurations with respect to
reflections at the equator simplifies the discussion of the
regular singular point. A careful numerical analysis shows 
that the solutions with horizon (bag of gold solutions) 
share the stability properties of the compact solutions.

The stability properties of purely magnetic, spherically symmetric,
static solutions to the EYM equations are determined by {\it two\/}, 
completely independent sets of perturbations. The investigation discussed above
applies to the even--parity fluctuations, 
which are also called ``gravitational'' perturbations, because they have no
flat spacetime analogues. Considering fluctuations of the non--magnetic
parts of the YM potential yields a second, orthogonal set of 
perturbations with {\it odd \/} parity. For the BK solitons and black holes,
these ``sphaleron--like'' perturbations give rise to additional $n$ 
unstable modes (see \cite{LM}, \cite{VBLS} 
and references therein). 
Adopting the method developed in \cite{VBLS}, we show in the last part
of this paper that this result remains true for both the compact
solutions {\it and} the solutions with horizon
found in \cite{VSLHB}.  

This article is organized as follows:
In the second section we give the basic equations for spherically 
symmetric EYM models and recall the most important features of the
compact solutions to these equations.

The third section is devoted to the derivation of the pulsation 
equations. We start by linearizing the YM equations within the 
purely magnetic ansatz. We then argue that -- by virtue of the 
Bianchi identity -- the $(00)$ and $(01)$ components of the Einstein equations 
give only rise to one independent linearized equation, which
has the form of a {\it constraint\/}. Finally, we linearize the partial
traces (with respect to $\gtiltens$ and the metric of $S^{2}$) of the 
Einstein equations, which yields two pulsation equations for the
metric fluctuations. The four equations obtained in this way can be
combined into a self--adjoint system. However, since no
particular gauge was fixed, the system still contains gauge modes, as
is also indicated by the form of the differential operator.

In the forth section we argue that there exists only one
gauge invariant quantity, $\zeta$, say, in the even parity sector. 
Unfortunately, the Sturm--Liouville equation for $\zeta$ has -- in 
addition to the origin -- a regular singular point at the equator. 
Since both branches of the fundamental system are analytic in the 
vicinity of the equator, it is not sufficient to consider only the 
solutions belonging to the unique self--adjoint extension of 
the fluctuation operator (i.e., the solutions for which $\zeta$
vanishes at the equator).

The reconstruction of the perturbations from the gauge invariant
quantity is presented in the fifth section. 
Using the globally regular {\it conformal} gauge and the general 
pulsation equations derived in the third section, we give explicit,
regular integral expressions for the metric and the matter fluctuations
in terms of $\zeta$. We also explain how to use the residual gauge 
freedom in order to fix the integration functions. 
The results of the numerical analysis are presented at the end of
this section.

An analytic discussion of the stability properties of the compact 
ground state solution 
($n=1$, $\Lambda = \Lambda_{\mbox{\scriptsize reg}}(1)$) is presented
in the sixth section. It turns out that the unstable mode corresponds
to spatially constant, conformal perturbations of the spacetime
metric. This ``breathing'' mode is also obtained by an alternative 
approach, which takes advantage of the special geometry of the ground 
state, in order to introduce the Bardeen potentials \cite{JB} used in 
cosmological perturbation theory 
(see, e.g., \cite{MFB}). We conclude
this section by deriving a ``dual'' Schr\"odinger equation for the
perturbations of the ground state. In contrast to the equation for
$\zeta$, the new pulsation equation for the ``supersymmetric partner''
of $\zeta$ is completely {\it regular\/} at the equator.  

In the last section of this paper we consider the perturbations with
odd parity. Again, the fluctuation equations can be reduced to a single 
Schr\"odinger type equation. Since the potential in this equation 
is unbounded at the zeroes of the YM amplitude $w$, we use the
residual gauge freedom to perform a supersymmetric transformation.
In this way, we obtain a pulsation equation with everywhere regular 
potential. We then demonstrate that the zero energy solution to this
equation has the same number of nodes as $w$. This eventually proves 
that both the compact and the bag of gold solutions have $n$ 
unstable modes in the ``sphaleron'' sector.

\section{Preliminaries}

In this paper we consider the question of
linear stability for the solutions of the
EYM equations with cosmological constant
presented in \cite{VSLHB}. 
Before we do so,
we shall briefly recall the basic notions
and equations used in \cite{VSLHB}. 
Since we are particularly interested in 
the globally regular, compact configurations, 
we also recall their basic features.

\subsection{Static Equations}

The metric of a spherically symmetric manifold $(M,\gtens)$ 
can be written in the form
\be
\gtens \, = \, \gtiltens \, + \, R^{2} \, \ghattens \, ,
\label{g-four}
\ee
where $\gtiltens$ is the metric on the pseudo--Riemannian
manifold $\tilde{M}$ and $\ghattens$ denotes the standard metric on 
$S^{2}$.
The Einstein tensor for the metric (\ref{g-four}) becomes
(see, e.g. \cite{BHS})
\bea
G_{ab} & = & \frac{2}{R}  \left[
\gtil_{ab} \daltil R - \nabtil_{a} \nabtil_{b} R \right] 
\, + \, \frac{1}{R^{2}} \gtil_{ab} \left[
(dR | dR) - 1 \right] \, ,
\label{G-ab} \\
G_{Ab} & = & 0 \, ,
\label{G-Ab} \\
G_{AB} & = & 
R^{2} \ghat_{AB} \left[ \frac{1}{R} \daltil R - \frac{1}{2} 
\tilde{\cal R} \right] \, ,
\label{G-AB}
\eea
where the quantities with a tilde
refer to $(\tilde{M},\gtiltens)$ and those with a hat to
$(S^{2},\ghattens)$.
(We use small and capital Latin letters for indices on 
$(\tilde{M},\gtiltens)$ and $(S^{2},\ghattens)$, respectively;
$a,\ b,\ c\ = 0,\ 1$ and $A,\ B,\ C\ = 2,\ 3$.) 
Without loss of generality, we shall often use the
diagonal parametrization 
\be
\gtiltens \, = \, - \, e^{2a(t,\rho)} \, dt^{2} \, + \, 
e^{2b(t,\rho)} \, d \rho^{2} 
\label{g-diag}
\ee
of the metric on $\tilde{M}$.
With respect to this, the d'Alembertian
of a function (e.g., $R$) and the Ricci scalar on 
$(\tilde{M},\gtiltens)$ become
(with $\dot{R} \equiv \partial R / \partial t$,
$R' \equiv \partial R / \partial \rho$)
\be
\daltil R \, = \, e^{-(a+b)} \left[ \, (e^{a-b} R')' - 
(e^{b-a} \dot{R} ) \, \dot{} \, \right] \, ,
\label{dAl}
\ee
and
\be
\tilde{\cal R} \, = \, -2\, e^{-(a+b)} \left[ \, (e^{a-b} a')' - 
(e^{b-a} \dot{b} ) \, \dot{} \, \right] \, ,
\label{R-til}
\ee
respectively.

A purely magnetic, spherically symmetric $SU(2)$ YM 
gauge potential is parametrized in terms of a scalar function
$w: \, \tilde{M} \rightarrow \RR$ (see, e.g. \cite{BK})
\be
A \, = \, (w -1) \, \left[
\, \hat{\tau}_{\varphi} \, d \vartheta \, - \, 
\hat{\tau}_{\vartheta} \, \sin \! \vartheta \, d \varphi \, \right] \, .
\label{YM-1form} 
\ee
The stress--energy tensor for $A$ has the components
\bea
8 \pi g^2 T_{ab} & = & \frac{1}{R^{2}} \left[
2 \, w,_{a} w,_{b} \, - \, \gtil_{ab} \, 
\left( \frac{1}{2} (dw | dw) + 
\frac{V(w)}{4 \, R^{2}} \right) \right] \, ,
\label{T-ab} \\
8 \pi g^2 T_{Ab} & = & 0 \, ,
\label{T-Ab} \\
8 \pi g^2 T_{AB} & = & \ghat_{AB} \, 
\frac{V(w)}{4 \, R^{2}} \, ,
\label{T-AB}
\eea
where $g^2$ is the gauge coupling constant and 
$V(w) \equiv (1-w^{2})^{2}$.

Using the notation
\be
R \, \equiv \exp(\mu) \, ,
\label{Rmu}
\ee
and the parametrization (\ref{g-diag}) of the metric
$\gtiltens$, the {\em static} field equations are
\bea
- \, e^{-2 b} \, \left[
\mu'' + \mu' \, (\mu' - a' - b' ) \right] & = &
\kappa \, e^{-2 b} \, \frac{w'^2}{R^2} \, ,
\label{s-0}\\
\frac{1}{R^2} \, - \, 
e^{-2 b} \, \left[
\mu'' + \mu' \, ( 2 \mu' + a' - b' ) \right]
& = &
\kappa \, \frac{V(w)}{2 \, R^4} \, + \, \Lambda \, ,
\label{s-1}\\
\; \;
\frac{1}{R^2} \, + \, 
e^{-2 b} \, \left[
a'' + a' \, ( a' - b' ) - \mu'^2 \right]
& = &
\kappa \, \frac{V(w)}{R^4} \, ,
\label{s-2}
\eea
and
\be
e^{-(a+b)} \, (e^{a-b} \, w')' \, = \, \frac{1}{4 \, R^{2}} \, V,_{w} \, .
\label{s-3}
\ee
Note that eqs. (\ref{s-0}), (\ref{s-1}) and (\ref{s-2})
are the $\frac{1}{2} (00+11)$,
$\frac{1}{2} (00-11)$ and
$\frac{1}{2} (00+11-22-33)$ components of the Einstein equations.
We also recall that the dimension--full coupling constant
$\kappa = 8 \pi G / g^2$ can be absorbed by using the dimensionless
quantities $R / \sqrt{\kappa}$,
$\rho / \sqrt{\kappa}$ and $\Lambda \kappa$ (see \cite{VSLHB}).
Hence, we use $\kappa = 1$ throughout.
(This corresponds to Johnstone Stoney units, already used in
1881 \cite{JS}.)

\subsection{Compact Solutions}

In the first six sections of 
this paper we are mainly interested in the stability
of the compact, regular solutions.
These are characterized in terms of the node number $n$
of the YM amplitude $w$ and the value of the cosmological constant, 
$\Lambda = \Lambda_{\mbox{\scriptsize{reg}}}(n)$. 
The lowest solution belongs to
$\Lambda = 3/(2\kappa)$ and can be given in closed form,
\bdm
a \, = \, 0 \, ,
\; \; \; \; 
b \, = \, 0 \, ,
\; \; \; \; 
R / \sqrt{\kappa} \, = \, \sin (\rho / \sqrt{\kappa}) \, ,
\edm
\be
w \, = \, \cos ( \rho / \sqrt{\kappa}) \, ,
\; \; \; \; \mbox{with} \; \, \Lambda = 3/(2\kappa) \, .
\label{comp-sol}
\ee
The solution describes the static Einstein Universe,
$I \! \! R \times S^{3}$, with
$T_{00} = \frac{3}{4} g^{-2} \frac{1}{8 \pi}$ and 
$T_{11} = T_{22} = T_{33} = \frac{1}{4} g^{-2} \frac{1}{8 \pi}$.
(For later use we also note that -- for $\kappa = 1$ -- we have
$\mu' = \cot \rho$,
$V(w) = \sin \! ^{4} \rho$, and
$V,_{w}/(4R^{2}) = \daltil w = -\cos \rho$.)

We recall that there are no solutions of the static 
equations for which $\mu'$ has more than one zero. (The first
non-vanishing derivative of $\mu$ is negative for every
$\rho$ with $\mu'(\rho) = 0$; see the expansions below.)
The unique value $\rho_{e}$ with $\mu'(\rho_{e}) = 0$
(and $R(\rho_{e}) \neq 0$) is therefore called the equator.
There are two families of solutions which are analytical in the 
vicinity of the equator. According to the behavior of the 
YM amplitude $w$, these solutions will be called odd
and even.

For the odd configurations,
$w(\rho_e) = 0$, $w'(\rho_{e}) \neq 0$, one 
finds with $x \equiv \rho - \rho_{e}$
(and $\kappa = 1$)
\bea
\mu' \, = \, -(\frac{w'_e}{R_{e}})^{2} \, x \, + \, {\cal O}(x^{3}) 
\, , \; \; \; & &
w' \, = \, w'_{e} \, + \, {\cal O}(x^{2}) \, ,
\nonumber \\
a \, = \, a_{e} \, + \, {\cal O}(x^{2})
\, , \; \; \; & &
b \, = \, b_{e} \, + \, {\cal O}(x^{2}) \, ,
\label{exp-odd}
\eea
where the parameters 
$w'_{e}$ and $R_{e}$ are subject to the constraint
equation
\bdm
\Lambda R_{e}^{\, 4} \, - \, R_{e}^{\, 2} \, + \, \frac{1}{2} \, = \, 
e^{-2b} \, (R_{e} \, w'_{e})^{2} \, .
\edm
The even solutions, for which 
$w(\rho_e) \neq 0$ and $w'(\rho_{e}) = 0$, we have the expansions
\bea
\mu' \, = \, -\frac{1}{3} (\frac{w''_e}{R_{e}})^{2} \, x^{3} 
\, + \, {\cal O}(x^{5}) 
\, , \; \; \; & &
w \, = \, w_{e} \, + \, {\cal O}(x^{2}) \, ,
\nonumber \\
a \, = \, a_{e} \, + \, {\cal O}(x^{2})
\, , \; \; \; & &
b \, = \, b_{e} \, + \, {\cal O}(x^{2}) \, ,
\label{exp-evn}
\eea
where $w''_e = e^{2b_e} V,_w(w_e)/(4 R_e^2)$. The parameters
$w_{e}$ and $R_{e}$ must fulfill the constraint
equation
\bdm
\Lambda R_{e}^{\, 4} \, - \, R_{e}^{\, 2} \, + \, \frac{V(w_{e})}{2} 
\, = \, 0 \, .
\edm

\section{Pulsation Equations}

We now linearize the field equations
on a static background; that is, we write
\bea
& & w(t,\rho) \, = \,  w(\rho) \, + \, \dw (t,\rho) \, ,
\label{lin-w}\\
& & \mu (t,\rho) \, = \, \mu (\rho) \, + \, \dmu (t,\rho) \, ,
\label{lin-mu}\\
& & \gtil_{ab} (t,\rho) \, = \, \gtil_{ab} (\rho) \, + 
\, \dgtil_{ab} (t,\rho) \, ,
\label{lin-gab}
\eea
and require that $w(\rho)$, $\mu(\rho)$ and $\gtil_{ab}(\rho)$ 
solve the static equations (\ref{s-0})--(\ref{s-3}).
Here and in the following, first order quantities 
are highlighted by boldface letters. 

For solutions {\it without\/} equator, 
the function $R$ is an admissible 
coordinate and, moreover, one may choose the
gauge $\dmu = 0$ to analyze stability properties.
The system of linearized field
equations then reduces to one pulsation equation 
(the linearized YM equation) 
and a set of constraint equations
(the linearized Einstein equations). 
By virtue of these constraints one can eliminate
the gravitational perturbations from the YM
equation, which yields a pulsation equation for 
$\dw$ alone. This technique was used to
investigate the stability properties of the
Bartnik--McKinnon solitons and the $SU(2)$ EYM
black holes with hair \cite{SZ}, \cite{ZS}. 
A detailed account of the procedure for a general class of
matter models was recently presented in \cite{BHS}.

The Schwarzschild coordinate $R$ is not suited to
describe spatially compact solutions.
In addition, the gauge $\dmu = 0$ is not regular at the
equator. The aim of the following analysis is to
generalize the procedure mentioned above, without
adopting the gauge $\dmu = 0$. We do so by deriving
a pulsation equation for a gauge invariant quantity
$\dzeta$, say, which basically reduces to $\dw$ in the 
Schwarzschild gauge, $\dmu = 0$. 
We then argue that there exists an everywhere regular gauge
for which the metric and matter perturbations
can be constructed explicitly from $\dzeta$.

We start by linearizing the relevant field equations
in a static background. In all what follows,
we use
\bea
\daltil \dw & = & 
e^{-a+b} \left( e^{a-b} \dw' \right)' \, - \, 
e^{-2a} \dddw \, ,
\nonumber \\
\daltil w & = &
e^{-a+b} \left( e^{a-b} w' \right)' \, ,
\nonumber
\eea
and similarly for all other first order
and static background quantities, respectively.

\subsection{The Yang--Mills Equation}

The dynamical YM equation reads
\be
\daltil \, w \, = \, \frac{1}{4 \, R^{2}} \, V,_{w} \, .
\label{YM-dyn}
\ee
For the diagonal metric (\ref{g-diag}) we obtain 
the following expression for the variation of the d'Alembertian 
on a static background:
\be
\delta \daltil w \, = \, \daltil \dw \, - \, 2 \daltil w 
\cdot \db
\, + \, (dw \, | \, d \dc) \, ,
\label{del-dal}
\ee
where 
\be
\dc \, \equiv \, \da \, - \, \db \, .
\label{def-c}
\ee
Variation of the r.h.s. of eq. (\ref{YM-dyn}) gives
$V,_{ww}/(4R^{2}) \dw - 2 \daltil w \cdot \dmu$, where we 
have used the static equation (\ref{s-3}) in the 
second term. The linearized YM equation 
now assumes the form
\be
\left[ \daltil - \frac{V,_{ww}}{4 \, R^{2}} \right]
\, \dw \, = \, 2\ \daltil w \, \cdot \, (\db - \dmu) \, - \,
(dw \, | \, d \dc) \, .
\label{YM-lin1}
\ee
For later use we also write this in terms of the
$4$--dimensional d'Alembertian, 
\bdm
\dal \, = \, \daltil \, + \, 2 \, (\, d \mu \, | \, d \, \; \, \cdot \; ) \, ,
\edm 
and the perturbation $\dx$,
\be
\dx \, \equiv \, \frac{1}{R} \, \dw \, .
\label{dx-def}
\ee
A short computation yields 
$\frac{1}{R} \daltil \dw = \dal \dx - (R \dal \frac{1}{R}) \dx$, 
and thus
\be
\left[ - \dal + (R \dal \frac{1}{R}) + 
\frac{V,_{ww}}{4 \, R^{2}} \right]
\, \dx \, = \, \frac{V,_{w}}{2 \, R^{3}} 
\, (\dmu - \db) \, + \, \frac{1}{R} \, (dw \, | \, d \dc) \, .
\label{YM-lin2}
\ee

\subsection{The Constraint Equation}

All first derivatives of the fields with respect to $t$ enter 
$G_{00}$ and $T_{00}$ only quadratically.
The corresponding linearized Einstein equation assumes 
therefore the form of a constraint equation, that is, 
it contains no time derivatives of the linearized quantities.
On the other hand, $G_{01}$ and $T_{01}$ already are
linear expressions in terms of time derivatives, 
implying that the corresponding linearized Einstein 
equation is a total derivative of a quantity $F$, say, 
with respect to time.
A lengthy computation then shows that the linearized
$(00)$--equation is the total spatial derivative of $F$.

This connection between the linearized $(00)$-- and 
$(01)$--equations is also established directly, as a
consequence of the Bianchi 
identity. In order to derive this result,
we consider the tensor
\be
E_{\mu \nu} \, = \, G_{\mu \nu} \, - \, 
8 \pi T_{\mu \nu} \, + \, \Lambda \, g_{\mu \nu} \, ,
\label{e-tens}
\ee
and linearize the contracted Bianchi identity
for $E_{\mu \nu}$.
Since the background equations imply that only
variations of $E_{\mu \nu}$ contribute, we have 
\be
0 \, = \, \frac{1}{\sqrt{-g}} \, \left(
\sqrt{-g} \, \delta E_{t}^{\; \mu} \right),_{\mu} \, - \, 
\Gamma^{\mu}_{\; \nu t} \, \delta  E_{\mu}^{\; \nu} \, .
\label{BI-lin}
\ee
Now using the facts that 
$\Gamma^{t}_{\; tt}$, $\Gamma^{t}_{\; ij}$ and $\Gamma^{i}_{\; tj}$
vanish in a {\it static\/} spacetime and that
$\Gamma^{t}_{\; it} = \frac{1}{2} g^{tt} g_{tt,i}$,
$\Gamma^{i}_{\; tt} = - \frac{1}{2} g^{ij} g_{tt,j}$, 
we find for the second term in the above expression
\bdm
\Gamma^{\mu}_{\; \nu t} \, \delta  E_{\mu}^{\; \nu} \, = \,
\frac{1}{2} g_{tt,j} \, \left[
g^{tt} \delta E^{j}_{\; t} - g^{ij} \delta E^{t}_{\; i} \right]
\, = \, 0 \, .
\edm
Here we have again used the fact that the tensor
$E^{j}_{\; t}$ vanishes identically (off shell) in a static
spacetime, implying that
$g^{tt} \delta E^{j}_{\; t} = 
g^{t \mu} \delta E^{j}_{\; \mu} = 
\delta (g^{t \mu} E^{j}_{\; \mu}) = 
\delta E^{j t}$ and 
$g^{ij} \delta E^{t}_{\; i} = 
\delta E^{t j}$.)
In the spherically symmetric case we therefore obtain the identity
\be
0 \, = \, \left( \sqrt{-g} \, \delta E_{t}^{\; t} \right) \dot{}
\, + \, \left( \sqrt{-g} \, \delta E_{t}^{\; \rho} \right)' \, .
\label{BI-lin-s}
\ee
Since, as we shall see in a moment,  
$(\sqrt{-g} \delta E_{t}^{\; \rho})$ is already the
time--derivative of a quantity $F$, say,
\bdm
\dot{F} \, = \,  \sqrt{-g} \, \delta E_{t}^{\; \rho} \, ,
\edm
we can integrate eq. (\ref{BI-lin-s})  with respect to
$t$. This shows that -- up to a 
integration function of $\rho$ -- the linearized
$(00)$--equation can be obtained from the linearized
$(01)$--equation, 
\bdm
F' \, = \, - \, \sqrt{-g} \, \delta E_{t}^{\; t} \, .
\edm

For a diagonal background metric and 
a gauge where the variation of the off--diagonal part
vanishes as well, we find from eq. (\ref{G-ab}) 
\bdm
\delta G_{t \rho} \, = \, -\frac{2}{R} \, \left[
(\dRdot)' \, - \, \Gamma^{t}_{\; \rho t} \, \dRdot \, - \,
R' \, \delta \Gamma^{\rho}_{\; \rho t} \right] \, .
\edm
Taking advantage of the fact that the 
background is static, this yields
\be
\delta G_{0}^{\; 1} \, = \, -2 \, 
\frac{\partial}{\partial t} \, \left[
e^{-(a+b)} \, \left(
\dmu' \, + \, (\mu' - a') \dmu \, - \, \mu' \db 
\right) \right] \, .
\label{G01-lin}
\ee
The variation of the corresponding component of the
stress--energy tensor is immediately found from
eq. (\ref{T-ab}),
\be
8 \pi \delta T_{0}^{\; 1} \, = \, 2 \, 
\frac{\partial}{\partial t} \, \left[
\frac{1}{R^2} \, e^{-(a+b)} 
\, w' \, \dw \right] \, .
\label{T01-lin}
\ee
Integrating the linearized $(01)$--equation
(\ref{G01-lin}) with respect to $t$ gives 
(up to a function of $\rho$)
\be
\dmu' \, + \, (\mu' - a') \dmu \, - \, \mu' \db 
\, + \, \frac{w'}{R} \dx \, = \, 0 \, , 
\label{constr-lin}
\ee
where, as earlier, $\dx \equiv \frac{1}{R} \dw$.
(Since the integration function has the form of an 
inhomogeneity, it can be set equal to zero by an 
appropriate choice of the initial conditions. 
Indeed, the linearized $(00)$--equation shows
that eq. (\ref{constr-lin}) is fixed up
to a constant.)
The arguments presented above now imply that the
linearized $(00)$--equation is the
derivative of this expression with respect to
$\rho$. Hence, eq. (\ref{constr-lin}) comprises
the information of {\it both\/}, the linearized
$(00)$-- and $(01)$--equations.

\subsection{Gravitational Pulsation Equations}

It remains to linearize the $(11)$--equation and the
$(AB)$--equations. Since we have already exhausted
the information from the $(0a)$ components, and since
$(M,\gtens)$ is spherically symmetric, it is
sufficient to consider the partial traces
$g^{ab} G_{ab} = \gtil^{ab} G_{ab}$ and
$g^{AB} G_{AB} = R^{-2} \ghat^{AB} G_{AB}$, and
to linearize the resulting equations
\bea
\dal \mu \, - \, \frac{1}{R^2} & = &
- \, \Lambda \, - \, \frac{V(w)}{2 \, R^4} \, , 
\label{dyn-1}\\
\dal \mu \, - \, (d \mu \, | \, d \mu) \, - \, 
\frac{1}{2} \tilde{\cal R} & = &
- \, \Lambda \, + \, \frac{V(w)}{2 \, R^4} \, . 
\label{dyn-2}
\eea
(Here we have also used
$\dal \mu = R^{-1} \daltil R + R^{-2}(dR|dR)$.)
With respect to the diagonal parametrization (\ref{g-diag})
of $\gtil$ we find
\bdm
\delta \dal \mu \, = \, \dal \dmu \, - \, 
2 \dal \mu \cdot \db \, + \, (d \mu \, | \, d[\dc + 2 \dmu] ) 
\edm
and
\bdm
-\frac{1}{2} \delta \tilde{\cal R} \, = \, 
\daltil \db \, - \, 2 \daltil a \cdot \db 
\, + \, e^{-(a+b)} ( e^{a-b} \dc')' \, + \,
(da \, | \, d \dc) \, .
\edm
Also noting that 
$\delta (d \mu | d \mu) = -2 \db (d \mu | d \mu) + 2 (d \mu | d \dmu)$
and using the static equation (\ref{s-2}) in the form 
$\daltil a - (d \mu | d \mu) = \frac{V}{R^4} - \frac{1}{R^2}$,
we find the linearized equations
\bea
\dal \dmu & = & 2 \dal \mu \cdot \db \, + \, 
2 (\frac{V}{R^4} - \frac{1}{R^2})
\, \dmu \, - \, \frac{V,_w}{2 R^3} \, \dx
\nonumber\\
& - & 2 \, (d \mu \, | \, d \dmu) \, - \, (d \mu \, | \, d \dc) \, ,
\label{lin-1}
\eea
and
\bea
\dal (\dmu + \db) & = & -2 \frac{V}{R^4} \, \dmu \, + \, 
2 (\frac{V}{R^4} - \frac{1}{R^2})
\, \db \, + \, \frac{V,_w}{2 R^3} \dx
\nonumber\\
& + & 2 \, d^{\dagger} \, (d \mu \, \db) \, + \, (d \mu - da \, | \, d \dc) 
\, - \, \lap \dc \, .
\label{lin-2} 
\eea
Here $d^{\dagger}$ denotes the co--derivative with respect to the spacetime
metric (for instance, $d^{\dagger} (d \mu \db)$ $=$ 
$\dal \mu \cdot \db + (d \mu | d \db)$), and
$\lap$ is the $4$--dimensional d'Alembertian {\it without\/} the 
time--derivative part,
\bdm
\lap \dc \, \equiv \, \frac{1}{\sqrt{-g}} (\sqrt{-g} \dc')' \, .
\edm
It is worth noticing that the operators 
$(d \mu \, | \, d \, \cdot)$ and $-d^{\dagger} (d \mu \, \cdot)$ are formally
adjoint to each other, that is, for arbitrary functions $g$ and $h$,
we have
\bdm
h \, {\cal D}_{d \mu} \, g \, - \, 
g \, {\cal D}^{\dagger}_{d \mu} \, h \, = \, 
d^{\dagger} \, (g \, h \, d \mu) \, ,
\edm
where
\bea
{\cal D}_{d \mu} \, g & \equiv & (d \mu \, | \, d g) \, ,
\nonumber \\
{\cal D}^{\dagger}_{d \mu} \, g & \equiv & 
- \, d^{\dagger} \, (d \mu \, g \, ) \, = \, - \,
\left[ (\dal \mu) + {\cal D}_{d \mu} \right]
\, g \, .
\label{D-defs}
\eea

\subsection{Symmetric Form of the Pulsation Equations}

We shall now write the system of linearized equations 
(\ref{YM-lin2}), (\ref{constr-lin}), (\ref{lin-1}), (\ref{lin-2})
in the form
\be
\Tmatrix \, \dal \, \dvec
\; = \; \Mmatrix \, \dvec \, ,
\label{symm-1}
\ee
where $\Tmatrix$ and $\Mmatrix$ are $4 \times 4$ matrices,
$\Mmatrix$ is formally self--adjoint, and $\dvec$ 
parametrizes the perturbations - suitably arranged in a
$4$--vector (see below). Using the operators defined above, the 
spatial derivative of the constraint equation
(\ref{constr-lin}) can be written as
\be
0 \, = \, - \,
{\cal D}_{d \mu}^{\dagger} \, \db \, + \, 
{\cal D}_{d \mu - da}^{\dagger} \, \dmu \, - \,
\lap \, \dmu \, + \,
[\frac{1}{R} {\cal D}_{d w}]^{\dagger} \, \dx \, ,
\label{constr-lin-der}
\ee
where $[\frac{1}{R} {\cal D}_{d w}]^{\dagger} \, \dx$
$=$ $-d^{\dagger} (\frac{1}{R} dw \, \dx)$.

The linearized equations 
(\ref{constr-lin-der}), (\ref{lin-1}), (\ref{lin-2})
and (\ref{YM-lin2}) can also be obtained from
variations of the effective action 
(expanded to second order in the fluctuations)
with respect to $\dc=\da-\db$, $\db$, $\dmu$ and $\dx=R^{-1}\dw$.
This implies that
\be
\dvec \; = \; (\, \dc \, , \, \db \, , \, \dmu \, , \, \dx \, ) \; .
\label{def-vec}
\ee
The system
(\ref{constr-lin-der}), (\ref{lin-1}), (\ref{lin-2}), (\ref{YM-lin2}) 
has now the desired form (\ref{symm-1}), with
\be
\Mmatrix \; = \; \Pmatrix \; + \; \Dmatrix \; ,
\label{symm-2}
\ee
where
$\Pmatrix$ is the symmetric potential matrix
\[ \Pmatrix \, = \,
\left(
\begin{array}{cccc}
0 & 0 & 0 & 0 \\
0 & 2 \, \dal \mu & 2[\frac{V}{R^4}-\frac{1}{R^2}] & -\frac{V,_w}{2 R^3} \\
0 & 2[\frac{V}{R^4}-\frac{1}{R^2}] & -2 \, \frac{V}{R^4} & \frac{V,_w}{2 R^3} \\
0 & -\frac{V,_w}{2 R^3} & \frac{V,_w}{2 R^3} & - R \dal \frac{1}{R} 
- \frac{V,_{ww}}{4 \, R^2}
\end{array}
\right)
\]
($-R \dal \frac{1}{R} = \dal \mu - (d \mu | d \mu)$), and
$\Dmatrix$ is the formally self--adjoint differential operator
\[ \Dmatrix \, = \,
\left(
\begin{array}{cccc}
0 & - {\cal D}^{\dagger}_{d \mu} 
& ( \, {\cal D}^{\dagger}_{d \mu - da} - \lap \, ) & (R^{-1} {\cal D}_{dw})^{\dagger} \\
- {\cal D}_{d \mu} & 0 & -2 \, {\cal D}_{d \mu} & 0 \\
( \, {\cal D}_{d \mu - da} - \lap \, ) & -2 \, {\cal D}^{\dagger}_{d \mu} & 0 & 0 \\
R^{-1} {\cal D}_{dw} & 0 & 0 & 0
\end{array}
\right) \; .
\]
The matrix T in front of the d'Alembertian is
\[ \Tmatrix \, = \,
\left(
\begin{array}{cccc}
0 & 0 & 0 & 0 \\
0 & 0 & 1 & 0 \\
0 & 1 & 1 & 0 \\
0 & 0 & 0 & -1 
\end{array}
\right) \; .
\]
Since the system contains one constraint equation,
the rank of $\Tmatrix$ is three. In addition,
$-\Tmatrix$ has only two positive eigenvalues,
reflecting the fact that the effective action
for the system is indefinite. In fact, the kinetic
term in the effective action is
\bdm
R^2 e^{b-a} \, \left[ \,
\frac{1}{R^2} \, (\dwdot)^2 \, + \, (\dbdot)^2 \, - \,
(\dmudot + \dbdot)^2 \, \right] \, .
\edm
Hence, the pulsation equations do not form a 
hyperbolic system in standard representation. 
It is therefore not clear how to make sense of 
the notion of stability on the basis of the 
above equations. 
We shall now present a way out of this difficulty, 
by arguing that the linearized system contains 
actually only one relevant dynamical degree of 
freedom. The aim is to find a 
gauge invariant generalization
of the elimination procedure used in the
Schwarzschild gauge ($\dmu = 0$).

\section{Gauge Invariant Formulation}

\subsection{Gauge Transformations}

Until now we have not fixed a gauge for the
spacetime metric (apart from 
the vanishing of the shift and its variation). 
In order to isolate the physical degrees of freedom it is, however,
crucial to get rid of the gauge modes. The metric
perturbations and the scalar YM amplitude $\dw$
transform according to $(L_{X} g)_{\mu \nu}$ and $L_{X} w$.
For a static background with diagonal metric, this implies   
\bea
\da & \longrightarrow & \da \; + a' \, f \; + \; \dot{g} \; , 
\nonumber \\
\db & \longrightarrow & \db \; + b' \, f \; + \; f' \; , 
\nonumber \\
\dmu & \longrightarrow & \dmu \; + \mu' \, f \; ,
\nonumber \\
\dw & \longrightarrow & \dw \; + w' \, f \; ,
\label{gauge-transf} 
\eea
with $f \equiv X^{\rho}$ and  
$g \equiv X^{t}$. 
Since we require that the shift and its perturbation vanish,
$f$ and $g$ are subject to 
\be
e^{2 \, b} \, \dot{f} \; = \; e^{2 \, a} \, g' \, .
\label{f-g-constr}
\ee
It is important to note the following: The
perturbation $\da$ enters the linearized equations only
via the combination $\da' - \db' = \dc'$.
Using eq. (\ref{f-g-constr}), one observes that
the gauge transformation for this quantity
does not involve the gauge freedom $g$:
\be
\dc' \, \longrightarrow \, \dc' \, + \, 2 (a'-b') f' 
\, + \, (a'' - b'') f \, - \, e^{2b} \daltil f \, .
\label{trans-dc}
\ee
Hence, the gauge transformations of the {\it relevant\/}
quantities -- i.e., the perturbations 
$\db$, $\dmu$, $\dw$ and $\dc'$ which enter the
field equations -- involve
only {\it one\/} degree of freedom, $f$.

\subsection{The Schwarzschild Gauge}

As a first application, we consider the gauge $\dR = 0$, i.e., 
$\dmu = 0$, which we shall also call the Schwarzschild gauge.
The gauge transformation 
which achieves $\dmu = 0$ is regular as long as $\mu' \neq 0$, 
that is, as long as $R$ is an admissible coordinate for the 
background solution. 
An advantage of this gauge is that the linearized gravitational
equation (\ref{lin-1}) becomes also a constraint equation. We therefore 
end up with the YM equation,
\bdm
\left[ \daltil - \frac{V,_{ww}}{4 \, R^{2}} \right] \, \dw \, = \, 
2\ \daltil w \cdot \db \, - \, e^{-2b} w' \, \dc' \, ,
\edm
and the constraint equations (\ref{constr-lin}) and (\ref{lin-2}),
\bea
\mu' \, \db & = & \frac{1}{R^2} \, w' \, \dw \, ,
\label{S-gauge-1} \\
e^{-2 b} \mu' \, \dc' & = &
- \frac{2}{R^2} \, \daltil w \cdot \dw \, + \, 
2 \dal \mu \cdot \db \, .
\label{S-gauge-2} 
\eea
Here we have also used the static background
YM equation. Hence, as is well known (see, e.g., \cite{BHS}),
the gravitational perturbations can be 
eliminated in the gauge $\dmu = 0$. The linearized
YM equation therefore assumes the form of a Schr\"odinger
equation for $\dw$,
\bdm
\left[ \daltil - \frac{V,_{ww}}{4 \, R^{2}} \right] \, \dw \, = \, 
\frac{2 w'}{\mu'^2 R^2} \, \left[
2 \mu' \daltil w \, - \, w' \dal \mu \right] \cdot \dw \, . 
\edm
Now using
$e^{2b} \dal \mu = \mu'' + (b' - a' + 2 \mu') \mu'$ and
$e^{2b} \daltil w = w'' + (b' - a') w'$, this pulsation 
equation for $\dw$ can be brought into the simple form 
\be
\daltil \, \dw \, = \, \left[ 
\frac{V,_{ww}}{4 \, R^{2}} \, + \, 
2 e^{-(a+b)} \, \left(
\frac{e^{a-b} w'^2}{R \, R'}
\right)' \, \right] \, \dw \, .
\label{YM-dw}
\ee
If $R$ can be chosen as a coordinate, then this 
equation is suited to discuss the perturbations
of the system. However, if, for instance, the
static hypersurfaces are topological $3$--spheres, 
then $R'$ has a zero, where the second potential term in
eq. (\ref{YM-dw}) becomes unbounded (see below).

\subsection{The Gauge Invariant Pulsation Equation}

In the previous section we have seen that
-- in a gauge where $\dmu$ vanishes --
the system of linearized equations can be reduced
to a single pulsation equation.
We shall now argue that this can always be achieved.
More precisely, we show that
one can find a pulsation equation 
which involves only a particular, {\it gauge invariant\/}
combination of the perturbations.
It will turn out to be convenient to write
eq. (\ref{YM-dw}) as an equation for $\mu' \dw$:
\be
\left[ \daltil - 2e^{-2b} \frac{\mu''}{\mu'}
\partial_{\rho} \right] (\mu' \dw) = \left[ 
\frac{V,_{ww}}{4 \, R^{2}} + 
2 e^{-(a+b)} \left(
\frac{e^{a-b} w'^2}{R \, R'}
\right)' - \mu' \daltil \frac{1}{\mu'}
\right] \cdot (\mu' \dw) .
\label{zeta-eq-1}
\ee

It is immediately observed from
the transformation laws (\ref{gauge-transf}) that   
\be
\dzeta \; = \; \mu' \, \dw \, - \, w' \, \dmu 
\label{g-in}
\ee
is a gauge invariant combination of the perturbations.
Since $\dzeta$ and $\mu' \dw$ are identical in the 
Schwarzschild gauge ($\dmu = 0$), 
and since eq. (\ref{zeta-eq-1}) was derived
within this gauge, we conclude that
eq. (\ref{zeta-eq-1}) {\it is\/} 
the pulsation equation for $\dzeta$.

Using the static equation (\ref{s-0}),
$R''/R' = a' + b' -w'^2/(RR')$,
the potential on the r.h.s. of the pulsation equation
(\ref{zeta-eq-1}) can also be cast in the form
\bdm
\left[ \,
\frac{V,_{ww}}{4 \, R^{2}} \, + \, 
2 \, \daltil \, (a + b - \ln |R'|) \, - \, 
\mu' \daltil \frac{1}{\mu'} \, \right] \, .
\edm
Combining the second and third terms in this bracket
yields the final result
\be
\left[ \daltil - 2 e^{-2b} \frac{\mu''}{\mu'} \partial_{\rho} \right] 
\, \dzeta \, = \, \left[ 
\frac{V,_{ww}}{4 \, R^{2}} \, + \, 
2 \, \daltil \, (a + b - \mu) \, - \, 
\frac{\daltil \mu'}{\mu'}
\right] \cdot \dzeta \, .
\label{zeta-eq-2}
\ee

Before we discuss this equation, let us note the following:
It is not hard to see that the transformation laws
(\ref{gauge-transf}) imply that
\be
\mu'^2 \, \db \, + \, 
(\mu'' - \mu' b') \, \dmu \, - \, 
\mu' \, \dmu' 
\ee
is gauge invariant as well. However, by virtue of the
constraint equation (\ref{constr-lin}) and the static 
background equation (\ref{s-0}),
one easily sees that this is equal to
$R^{-2} w' \dzeta$.
In fact, the circumstance that we are able to
reconstruct all perturbations from $\dzeta$, and
the fact that the gauge transformations for the 
relevant quantities contain only one free function, $f$,
strongly suggest that the fluctuation problem
involves exactly one relevant gauge invariant quantity.

\subsection{The Sturm--Liouville Equation}

Let us now consider the pulsation equation (\ref{zeta-eq-2}) for 
the gauge invariant quantity $\dzeta$.
As expected, this is a Sturm--Liouville equation. 
Indeed, if we introduce the operator   
\be
\Aop \, = \, 
e^{a-b}{\mu'}^2\left[-{\partial_\rho}\,
\frac{\e^{a-b}}{{\mu'}^2}\,{\partial_\rho}\; 
\;+\;q(\rho)\right]
\label{brozeta_1}
\ee
with potential
\be
q(\rho)= \,\frac{\e^{a+b}}{{\mu'}^2}\left[ 
\frac{V,_{ww}}{4 \, R^{2}} \, + \, 
2 \, \daltil \, (a + b - \mu) \, - \, 
\frac{\daltil \mu'}{\mu'}\right]\;,
\ee
then eq. (\ref{zeta-eq-2}) becomes
\be
\ddot{\dzeta} \, + \, \Aop \,\dzeta \, = \, 0\;.
\label{brozeta_2}
\ee

Here we would like to comment 
on a subtlety arising in the study of this
equation. A standard approach is to first consider
eq. (\ref{brozeta_2}) as an ordinary differential 
equation in some $L^2$--space.
In a second step one then discusses under which conditions a 
Hilbert space solution can be identified with a {\it strict\/} 
solution 
in the ordinary sense.
In our case, the differential equation 
corresponding to the operator $\Aop$ 
obviously has singularities at the origin ($\rho = 0$) 
and at the equator ($\rho =\rho_e$). 
(In fact, both points are {\it regular\/}
{\it singular\/} points; see below.)
It is therefore natural to consider $\Aop$ as an operator 
on the Hilbert space L$^2(I,d\eta)$, where $I$ is the open 
interval $(0,\rho_e)$ and $d\eta$ 
denotes the measure $e^{a-b} \mu'^{-2} d\rho$. 
Then $\Aop$ is symmetric on a dense domain 
and, as we will show shortly,
for both boundary points the {\it limit point\/} case 
(in the sense of Weyl) occurs. 
Now Weyl's alternative (see, e.g., \cite{Book}) 
implies that there is precisely one self-adjoint realization of $A$. 
Moreover, this realization is essentially self-adjoint on
$$
{\cal D}=\{\,u\in 
\mbox{C}^2(I)\cap \mbox{L}^2(I,d\mu)\mid Au\in \mbox{L}^2(I,d\mu)\;\}\;.
$$
In particular, all functions in $\cal D$ 
vanish at the origin {\em and} at the equator. 
On the other hand, as we shall demonstrate later,
there exist also analytic solutions 
with $\dzeta(\rho_e)\neq 0$ which give rise to 
physically acceptable metric perturbations.
By adopting the functional analytic approach, we
would thus loose half of the interesting modes.

In order to establish that the equator is
a {\em regular\/} {\em singular\/} point, 
we note that the factors
in front of $\dzeta'$ and $\dzeta$ diverge not stronger than
$x^{-1}$ and $x^{-2}$, respectively.
More precisely, we have
\be
\dzeta'' \, - \, \left[
\, 2\, \frac{\mu''}{\mu'} \, + \, {\cal O}(x) \, \right]
\, \dzeta' \, = \, \left[
\, - \frac{\mu'''}{\mu'} \, + \, {\cal O}(1) \, \right]
\, \dzeta \, ,
\label{zeta-equator}
\ee
where
$\mu' = {\cal O}(x^{n})$, with $n=1$ for odd and $n=3$ for 
even background configurations.
The indicial polynomial for the above equation is
\bdm
r \, (r-1) \, - \, 2 \, r \, n \, + \, n \, (n-1) \, = \, 0 \, ,
\edm
which has the roots 
\[
r_\pm \, = \, \frac{1}{2} \, \left[ 
1 + 2 n \, \pm \, \sqrt{1 + 8 n} \right] \, 
\, = \, \, \left\{
\begin{array}{ll}
3, \, 0 & \mbox{odd case,} \\
6, \, 1 & \mbox{even case.}
\end{array}
\right. 
\]
Together with the symmetry 
of the background solution, this 
implies that the fundamental system of 
eq. (\ref{zeta-eq-2}) is analytic near the equator. 
For $r_\pm = 3,0$ this has an expansion of the form
\be
\dzeta_{+} \, = \, x^{3}\,\sum_{j=0}^{\infty} \, c_{j}\, x^{2j} \, ,
\qquad
\dzeta_{-} \, = \, \sum_{j=0}^{\infty} \,
k_{j}\, x^{2j} \,, 
\label{fund-odd}
\ee
with $c_{0} = k_{0} = 1$. 
For $r_\pm = 6,1$ one has
\be
\dzeta_{+} \, = \, x^6\sum_{j=0}^{\infty}c_{j}\, x^{2j} \,,
\qquad
\dzeta_{-} \, = \, x\,\sum_{j=0}^{\infty} \,
k_{j}\, x^{2j} \, ,
\label{fund-even}
\ee
with $c_{0} = k_{0} = 1$.

Let us also discuss the origin. To leading order, 
eq. (\ref{zeta-eq-2}) reads
\be
\dzeta'' \, + \, \left[
\, 2 \,\frac{1}{\rho} \, + \, {\cal O}(1) \, \right]
\, \dzeta' \, = \, \left[
\, \frac{2}{\rho^{2}} \, + \, {\cal O}(\rho^{-1}) \, \right]
\, \dzeta \, .
\label{zeta-eq-3}
\ee
With a similar reasoning as above, we now obtain the fundamental
solutions
\be
\dzeta_{+} \, = \, \rho\sum_{j=0}^{\infty}c_{j}\, \rho^{2j} \,,
\qquad
\dzeta_{-} \, = \, \frac{1}{\rho^2}\,\sum_{j=0}^{\infty} \,
k_{j}\, \rho^{2j} \,.
\label{poleexp}
\ee

It is now easy to verify that 
both boundary points of the operator $\Aop$
belong to the limit point case.
For $\rho = \rho_e$ one has to show that the 
differential equation $\Aop u=0$ has a solution for which
\be
\int_{\rho_e-\varepsilon}^{\rho_e}|u|^2\,d\eta \, = \, \infty
\ee
for some $\varepsilon>0$ (see, e.g., \cite{Book}). 
(For $\rho = 0$ a solution with the corresponding 
property in the vicinity of the origin has to be found.) 
Using the behavior of the measure $d\eta=\e^{b-a} \mu'^{-2} d\rho$ 
at the singular points and the 
above expansions for the fundamental systems establishes 
that for both $\rho = 0$ and 
$\rho = \rho_{e}$, $\Aop$ is in the limit point case.

\section{Reconstruction of the Perturbations}

It remains to show that every solution of the
pulsation equation for $\dzeta$ gives rise to 
a regular set $(\dw, \dmu, \db, \dc')$ of perturbations.
One may try to achieve this in a gauge where 
the relevant equations reduce to a system 
of algebraic equations for $(\dw, \dmu, \db, \dc')$  
in terms of $\dzeta$. The two gauges of this
kind are the Schwarzschild gauge,
$\dmu = 0$, and the gauge $\dw = 0$. 
However, none of these gauges
yields a globally regular description for
solutions which posess an equator, which demonstrates
that this method does not always work.

A better way is to make use of a suitable gauge
for which the perturbations can be obtained from a system of
{\it ordinary\/} differential equations.
This is the case for any
gauge of the form $\db = constant \cdot \dmu$,
because one can then eliminate 
the d'Alembertians of $\dmu$ and $\db$
form the pulsation equations (\ref{lin-1}) and (\ref{lin-2}).
In this way one obtains an additional {\it ordinary\/} 
differential equation for $\dc$. There are two representatives 
within this class which yield globally regular
expressions for the perturbations 
in terms of $\dzeta$. These are $\db = 0$ and 
the {\it conformal\/} gauge, $\db = \dmu$.

\subsection{Coverings with two Gauges}

Before we shall work in the conformal gauge, let us briefly
consider the two gauges
for which the perturbations are algebraic expressions
in terms of $\dzeta$.
Off the equator, $\mu'$ does not vanish and the 
Schwarzschild gauge, $\dmu = 0$, is 
therefore perfectly regular.
(The gauge transformation 
$\dmu \rightarrow 0 = \dmu + \mu' f$
is regular for $\mu' \neq 0$.)
Using eqs. (\ref{S-gauge-1}) and (\ref{S-gauge-2}) one easily
finds for the perturbations in terms of $\dzeta$
\bea
\dmu \, = \, 0 \, , 
\; \; \; \; & &
\dw \, = \, \dfrac \, ,
\label{e-1}\\
\db \, = \, \frac{w'}{R R'} \, \dw \, ,
\; \; \; \; & &
\dc' \, = \, - 2 \left( 
\frac{w'}{R R'} \right)' \, \dw \, .
\label{e-2}
\eea
If the background solution has no equator, then the
Schwarzschild gauge is everywhere regular and
the above formulae can be used to compute
the perturbations in the entire spacetime. 
If, however, we are considering solutions
where $R$ assumes a local maximum, we need
another gauge in the vicinity of the equator.
The odd case, $w'_e \neq 0$,
is very simple to handle, since one can 
then choose the gauge $\dw = 0$ and express
$\dmu$ in terms of the gauge invariant quantity
$\dzeta$. Then using the constraint equation
(\ref{constr-lin}), we obtain $\db$ in terms of $\dzeta$.
Eventually, taking advantage of the linearized 
YM equation (\ref{YM-lin1})
in the gauge $\dw = 0$, we also obtain 
$\dc'$ in terms of $\dzeta$:
\bea
\dw \, = \, 0 \, , 
\; \; \; \; & &
\dmu \, = \, - \, \frac{\dzeta}{w'} \, ,
\label{e-3}\\
\db \, = \, \frac{1}{R e^{-a}} \,
\frac{(R e^{-a} \, \dmu)'}{\mu'} \, ,
\; \; \; \; & &
\dc' \, = \, 2 e^{2b} \, \frac{\daltil w}{w' e^{-a}} \,
\frac{(e^{-a} \, \dmu)'}{\mu'} \, .
\label{e-4}
\eea
It is clear that $\dmu$ remains finite since,
by assumption, $w'$ does not vanish in a sufficiently
small neighborhood of the equator. Moreover,
we have $\dmu = constant + {\cal O}(x^2)$,
since both $\dzeta$ and $w'$ are even.
Since the background function $R e^{-a}$ is even
as well, we conclude that the numerators in the
expressions for $\db$ and $\dc'$ are of ${\cal O}(x)$.
Also using $\mu' = {\cal O}(x)$ shows that
both $\db$ and $\dc'$ are well behaved in the 
vicinity of the equator.

\subsection{A Globally Regular Gauge}

The gauge $\dw = 0$ is not suited to compute the 
perturbations in the even case, $w'_{e} = 0$,
since the expressions in eq. (\ref{e-4}) diverge
if $\mu' = {\cal O}(x^{3})$.
We now show that there exists a
{\em regular} gauge which gives rise to analytic
expressions in the vicinity of the
equator for {\em both} odd and even background configurations.
Moreover, this gauge is also regular at the
poles, where it yields analytic expressions for the perturbations
as well. It is therefore possible to cover the 
entire spacetime using just {\it one\/} gauge.

We consider {\it conformal\/} perturbations of the
spatial part of the metric, that is - 
without loss of generality - we choose the gauge
\be
\db \; = \; \dmu \, .
\label{db-gauge}
\ee
(We also note that the gauge transformation between
a globally regular gauge and the conformal gauge is everywhere 
regular.)

We have already mentioned that -- by virtue of 
any algebraic relation between 
$\db$ and $\dmu$ -- it is possible to eliminate 
the d'Alembertians from the pulsation equations (\ref{lin-1}) 
and (\ref{lin-2}), which therefore provides 
an additional {\it ordinary\/} differential equation for $\dc'$.
For a given solution $\dzeta$ of the gauge invariant
pulsation equation one can then
reconstruct all perturbations from 
$\dzeta$ by solving a system of
{\it constraint\/}  equations.
The relevant equations are 
the pulsation equation (\ref{zeta-eq-2}) for $\dzeta$,
the algebraic relations (\ref{db-gauge}) and (\ref{g-in}), the
constraint (\ref{constr-lin}) and the new constraint
(\ref{sy-4}), obtained from eqs. 
(\ref{lin-1}) and (\ref{lin-2}):
\be
\left[ \daltil - 2 e^{-2b} \frac{\mu''}{\mu'} \partial_{\rho} \right] 
\, \dzeta \, = \, \left[ 
\frac{V,_{ww}}{4 \, R^{2}} \, + \, 
2 \, \daltil \, (a + b - \mu) \, - \, 
\frac{\daltil \mu'}{\mu'}
\right] \cdot \dzeta \; ,
\label{sy-1}
\ee
\be
\mu' \, \dw \, - \, w' \, \dmu \; = \; \dzeta \; ,
\label{sy-2}
\ee
\be
\dmu' \, - \, a' \, \dmu \, + \, \frac{w'}{R^2} \, \dw \; = \; 0 \; ,
\label{sy-3}
\ee
\be
\left( \frac{e^{2a-b}}{R} \, \dc' \right)' \; = \; 
{\cal W} \, \dw \, + \, {\cal M} \, \dmu \; ,
\label{sy-4}
\ee
where
\bea
{\cal W} & \equiv & 6 \, R^{-3} e^{2a+b} \, [\daltil w 
\, - \, (dw | d \mu)] \, ,
\nonumber \\
{\cal M} & \equiv & -2 \, R^{-1} e^{2a+b} \, [\daltil (\mu + 2a) 
\, - \, 3 (da | d \mu) \, + \, \frac{1}{R^2}] \, .
\label{sy-5}
\eea

It is worth noticing 
that the gauge transformation $f$ which
achieves $\db -  \dmu \rightarrow 0$ is everywhere regular,
in particular at the equator ($R' = 0$) and at the poles ($R = 0$). 
In fact, one easily finds from eqs. (\ref{gauge-transf}) 
\be
f \, = \, - \, R \, e^{-b} \, \int R^{-1} e^{b} \, (\db - \dmu) \, .
\label{k-gtr}
\ee
The regularity at the equator is manifest, whereas the 
analyticity at the poles follows from the fact that
$\db - \dmu = {\cal O}(R)$. (Use the Schwarzschild gauge $\dmu = 0$
for which $\db = {\cal O}(R)$ and the fact that the transformation from
the Schwarzschild to the conformal gauge is ${\cal O}(R^2)$.) 

>From a given solution $\dzeta$ of the gauge invariant
pulsation equation we now obtain $\dmu$ and $\dw$ 
from the definition (\ref{sy-2}) of $\dzeta$
and the constraint (\ref{sy-3}):
\bea
\dmu & = & \dmu(\rho_0,t) \,
 - \, R' \, e^{-b} 
\int_{\rho_0}^{\rho}  
\frac{e^{b}}{R} \,
\frac{w'}{R'^{2}} \, \dzeta \, ,
\label{k-dmu}\\
\dw & = & \dw(\rho_0,t) \, + \,
w' \, R \, e^{-b} \, 
\int_{\rho_0}^{\rho}  
\frac{e^{a+b}}{R'}
\left( e^{-a} \frac{1}{w'} \, \dzeta \right)' \, ,
\label{k-dw}
\eea
where we have also used the background equation (\ref{s-0}).
Integrating the additional constraint equation (\ref{sy-4})
finally gives $\dc'$: 
\be
\dc' \, = \, \dc'(\rho_0,t) \, + \,
R \, e^{b-2a} \, 
\int_{\rho_0}^{\rho} \,
\left[ \, {\cal W} \, \dw \, + \, {\cal M} \, \dmu \, \right] \, .
\label{k-dc}
\ee
The time dependent quantities 
$\dmu(\rho_0,t)$, $\dw(\rho_0,t)$ and $\dc'(\rho_0,t)$
are not independent. The fact that they
are subject to the YM equation (\ref{YM-lin1}),
together with the residual
gauge freedom, implies that they are 
uniquely determined by $\dzeta$
(see the next section).

The perturbations obtained from eqs. (\ref{k-dmu})-(\ref{k-dc})
are analytic in the vicinity of the equator
for both odd and even background configurations:
In the even case, where 
$w' = {\cal O}(x)$, $R' = {\cal O}(x^{3})$ and
$\dzeta = {\cal O}(x)$, the integrands in eq. (\ref{k-dmu}) 
and (\ref{k-dw}) diverge like 
${\cal O}(x^{-4})$ and ${\cal O}(x^{-2})$, respectively. 
Hence, $\dmu = {\cal O}(x^{3}) \int {\cal O}(x^{-4}) = {\cal O}(1)$
and $\dw = {\cal O}(x) \int {\cal O}(x^{-2}) = {\cal O}(1)$.
In the odd case, where
$w' = {\cal O}(1)$, $R' = {\cal O}(x)$ and
$\dzeta = {\cal O}(1)$, one has
$\dmu = {\cal O}(x) \int {\cal O}(x^{-2}) = {\cal O}(1)$
and $\dw = {\cal O}(1) \int {\cal O}(1) = {\cal O}(x)$.

The above formulae give also rise to analytic expressions
in the vicinity of the poles. Using the behavior of the
background solution for $R \rightarrow 0$ we have
already concluded from eq. (\ref{zeta-eq-3}) 
that $\dzeta = {\cal O}(R)$.
Now using $w' = {\cal O}(R)$ shows that 
$\dmu = {\cal O}(1)$ and $\dw = {\cal O}(R)$.
Since the terms of ${\cal O}(R^{-3})$
in ${\cal M}$ cancel, we have
${\cal M} \dmu = {\cal O}(R^{-2})$ and
${\cal W} \dw = {\cal O}(R^{-2})$. Hence,
$\dc' = {\cal O}(R) \int {\cal O}(\frac{1}{R^2}) 
= {\cal O}(1)$.

In conclusion, we have shown that the gauge $\db = \dmu$ 
gives everywhere regular expressions
for the entire set of perturbations obtained from
the gauge invariant quantity $\dzeta$.
The same also holds for the gauge $\db = 0$, which turns
out to be particularly suited for the numerical
investigation (see Sect. 5.4).

\subsection{The Residual Gauge Freedom}

It remains to discuss the 
implications of the residual
gauge freedom. The transformation laws 
(\ref{gauge-transf}) imply
\bdm
(\db - \dmu) \, \longrightarrow
(\db - \dmu) \, + \, (b' - \mu') \, f \, + \, f' \, .
\edm
The conformal gauge 
$\db - \dmu = 0$ is therefore invariant under transformations
of the form $f(\rho,t) = R e^{-b} \, \phi(t)$, where
$\phi(t)$ is an arbitrary function which does not
depend on $\rho$. The remaining gauge
freedom is thus
\bea
\dmu \, \longrightarrow & & \dmu \; + R' \, e^{-b} \, \phi(t) \; ,
\nonumber \\
\dw \, \longrightarrow & &\dw \; + w' \, R \, e^{-b} \, \phi(t) \; ,
\nonumber \\
\dc' \, \longrightarrow & & \dc' \; + \,
R \, e^{b - 2a} \, \ddot{\phi}(t) \, - \,
\left[ e^{a-b} \left( R e^{-a} \right)' \right]' \, \phi(t)
\nonumber \\
& & = \, \dc' \; + \,
R \, e^{b - 2a} \, 
[ \, \ddot{\phi} \, - \, k(\rho) \, \cdot \, \phi  \, ] \, ,
\label{dc-bis}
\eea
with 
\be
k(\rho) \, \equiv \, R^{-1} e^{2a-b} \, \left[ e^{a-b} \left(
R \, e^{-a} \right)' \right]' \, .
\label{def-k}
\ee
 
Comparing these expressions with the integral formulas
(\ref{k-dmu})-(\ref{k-dc}) shows that
contributions to $\dmu$ of the
form $R' e^{-b} \alpha(t)$ can be compensated by
a gauge transformation with $\phi(t) = -\alpha(t)$.
(The same holds for terms of the
form $w' R e^{-b} \beta(t)$ in the expression for $\dw$.)
Hence, by virtue of the residual gauge freedom,
$\dmu$ is uniquely determined by $\dzeta$.
Since we already know from the integral expressions
that $\dw$ and $\dc'$ are regular, we can compute
these perturbations from the definition
(\ref{g-in}) of $\dzeta$ and the
linearized YM equation. 
This shows that -- once the ambiguity in
$\dmu$ has been removed by the residual gauge freedom -- all 
perturbations are uniquely determined in terms of $\dzeta$.

\begin{figure}
\epsfxsize=10cm
\centerline{\epsffile{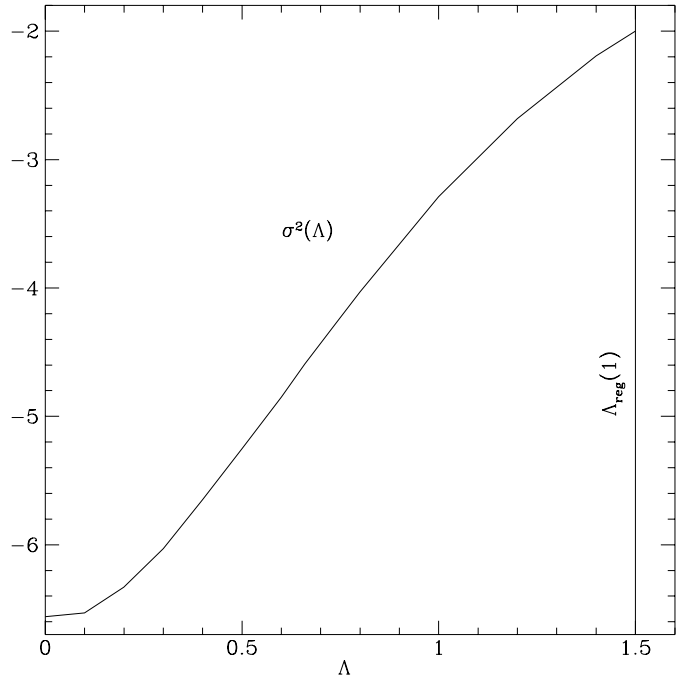}}
\caption{Bound state energy versus cosmological constant 
for the $n=1$ EYM solutions.}
\label{Fig. 1}
\end{figure}

\subsection{Numerical Results}

As already mentioned, the gauge $\db = 0$ is particularly
suited for the numerical analysis. Assuming a harmonic
time--dependence $e^{i \sigma t}$ for the fluctuations, the
YM equation (\ref{YM-lin1}) and the constraint (\ref{constr-lin})
become (with $\dc = \da$)
\be
\left[
\frac{\partial}{\partial \rho} Q \frac{\partial}{\partial \rho} 
\, - \, \frac{V,_{ww}}{4 R^{2}} 
\right] \, \dw \, + \, 
(Qw')' \, \dmu \, + \, (Qw') \, \da' \, = \, 
- \, Q^{-1} \sigma^{2} \, \dw \, ,
\label{num-1}
\ee
\be
\dmu' \, - \, \frac{w'}{R^{2}} \, \dw \, + \, 
(\mu' - a') \, \dmu \, = \, 0 \, .
\label{num-2}
\ee
Here we have also used the background gauge 
$e^{2a} = e^{-2b} \equiv Q(\rho)$. Eliminating
the d'Alem\-ber\-ti\-ans from 
eqs. (\ref{lin-1}) and (\ref{lin-2}), we obtain
the additional constraint equation 
\be
Q^{-1/2} \left(Q^{3/2} \da' \right)' \, - \, 
\frac{V,_{w}}{R^{4}} \, \dw \, - \, 
2 e^{-2b} \mu' \, \dmu' \, + \, 
\frac{2}{R^{4}} (2 V - R^{2}) \, \dmu \, = \, 0 \, .
\label{num-3}
\ee
The main advantage of the gauge $\db = 0$ is that the
above equations are manifestly regular at the critical
points $R' = 0$. They can therefore be used to determine the
spectrum for any background solution with $n$ nodes
and cosmological constant $\Lambda(n) \in [0 , \Lregn]$.

The numerical analysis shows that there exists exactly
one boundstate (in the gravitational sector 
under consideration) for each solution with $n = 1$ and
$\Lambda(1) \in [0 , \Lambda_{\mbox{\scriptsize reg}}(1)]$.
The energy of the unstable mode versus the cosmological constant
is shown in Fig. 1. For $\Lambda = 0$ we obtain the 
boundstate belonging to the lowest ($n = 1$) BK solution
(in the normalization $Q(0) =1$, $\kappa  =1$).
In the other limiting case, that is for $\Lambda(1) = 3/2$,
we find $\sigma^{2} = -2$, which is the eigenvalue
belonging to the unstable mode of the lowest compact solution 
(see also the next section).

A careful numerical investigation 
for the few lowest solutions strongly suggests that
there exist $n$ unstable modes (in the gravitational sector)
for any background solution with $n$ zeroes of $w$.

\section{Instability of the $n=1$ Compact Solution}

\subsection{Gauge Invariant Approach}

The stability analysis of the ground state solution,
\be
a \, = b \, = 0 \, , \; \; \; \; \; 
R \, = \, \sin \rho \, , \; \; \; \; \; 
w \, = \, \cos \rho \, ,
\label{ground-0}
\ee
can be performed analytically.
Using $\daltil \dzeta = - \dddzeta + \dzeta''$,
$\daltil \mu = \mu'' =-1 / \sin^{2} \! \rho$,
$\frac{1}{\mu'} \daltil \mu' = \mu'''/\mu' = 2 / \sin^{2} \! \rho$
and 
$(V,_{ww}) / (4 R^{2}) = (3 \cos^{2} \! \rho - 1) / \sin^{2} \! \rho$,
one finds from eq. (\ref{zeta-eq-2})
\be
- \, \dddzeta \, + \, \dzeta'' \, + \, \frac{2}{\sin \rho \, \cos \rho}
\, \dzeta' \, = \, 
\frac{3 \cos^{2} \! \rho \, - \, 1}{\sin^{2} \! \rho} \, \dzeta \, .
\label{ground-1}
\ee
The solutions of this equation of the form
$\dzeta(t, \rho) = \dzeta_{(k)}(\rho) e^{i \, \sigma_{k} t}$ are
\be
\dzeta_{(k)}(\rho) \, = \, \cos^{2} \! \rho
\, \left( \frac{\sin (k \rho)}{\sin \rho \, \cos \rho}
\right)' \, ,
\label{ground-2}
\ee
with the spectrum 
\be
\sigma_{k}^{2} \, = \, k^{2} \, - \, 3 \, , \; \; \; 
\mbox{where} \; k \, = \, 1, \, 2, \, \dots \, . 
\label{ground-3}
\ee
In the conformal gauge, $\db = \dmu$,  
we now obtain $\dmu$ from eq. (\ref{k-dmu}),
\be
\dmu \, = \, \cos \rho \, \int^{\rho} \frac{1}{\cos^{2} \! \rho} \, 
\dzeta \, \, = \, \frac{\sin (k \rho)}{\sin \rho} \, + \,
\cos \rho \cdot \alpha(t) \, ,
\label{ground-4}
\ee
where $\alpha(t)$ is a free function. 
In order to obtain $\dw$, we can either use
eq. (\ref{k-dw}) or the definition of $\dzeta$ or -- most easily -- the 
constraint (\ref{sy-3}), $\dw = \sin \rho \, \dmu'$:
\be
\dw \, = \, \sin \rho \left(
\frac{\sin (k \rho)}{\sin \rho} \right)' \, - \,
\sin^{2} \! \rho \cdot \alpha(t) \, .
\label{ground-5}
\ee

Finally, we have to evaluate the integral
(\ref{k-dc}) for $dc'$.
This is, however, trivial since 
the background quantities ${\cal W}$ and
${\cal M}$ defined in eq. (\ref{sy-5}) 
vanish for the ground state solution (\ref{ground-0}):
\bdm
{\cal W} = \frac{6}{R^{3}}(w'' - w' \mu') = 0 \, ,
\; \; \; \; 
{\cal M} = -\frac{2}{R}(\mu'' + \frac{1}{R^{2}}) = 0 \, .
\edm
We therefore conclude from eq. (\ref{k-dc}) that
$(\dc' / \sin \rho)' = 0$, that is, 
\be
\dc' \, = \, \sin \rho \, \cdot \, \gamma(t) \, .
\label{ground-6}
\ee
It is clear that $\gamma(t)$ is not an independent
free function, since $dc'$ is uniquely determined 
from $\dw$ via the linearized YM equation 
(\ref{YM-lin1}). Using the above solution for $\dw$ we find
\bdm
\left[ \daltil - \frac{V,_{ww}}{4 R^{2}} \right]
(- \sin^{2} \! \rho \cdot \alpha) \, = \, 
\left[ \ddot{\alpha} - \alpha ( \frac{\partial^{2}}{\partial \rho^{2}}
- \frac{3 \cos^{2} \! \rho -1}{\sin^{2} \! \rho} ) \right] 
\sin^{2} \! \rho
\edm
\bdm 
= \, (\ddot{\alpha} \, + \, \alpha ) \, \sin^{2} \! \rho \, = \, 
-w' \, dc' \, = \, \sin \rho \, \dc' \, ,
\edm
which shows that
\be
\gamma \, = \, \ddot{\alpha} \, + \, \alpha \, . 
\label{ga-al}
\ee

Eventually, we can use the remaining gauge freedom
to get rid of the free function $\alpha(t)$:
Considering transformations of the form
$f = R e^{-b} \phi(t) = \sin \rho \cdot \phi(t)$ with $\phi(t) = - \alpha(t)$
yields
\bea
\dmu \, \longrightarrow & & \dmu \; - \, \cos \rho \cdot \alpha(t) \; ,
\nonumber \\
\dw \, \longrightarrow & &\dw \; + \sin^{2} \! \rho \cdot \alpha(t) \; ,
\nonumber \\
\dc' \, \longrightarrow & & \dc' \; - \, \sin \rho \cdot 
(\ddot{\alpha} + \alpha) \, = \, \dc' \; - \, \sin \rho \cdot 
\gamma(t) \, .
\label{conf-gt}
\eea
This implies that - for the ground state -
it is consistent to consider conformal
perturbations of the {\it entire\/} spacetime metric
\bdm
\da \, = \, \db \, = \, \dmu \, = \frac{\sin (k \rho)}{\sin \rho}
\; e^{i \, \sigma_{k} \, t} \, ,
\edm
\be
\dw \, = \, \dmu' \; \sin \rho 
\; e^{i \, \sigma_{k} \, t} \, .
\label{ground-7}
\ee
It is also worth noticing that the residual gauge freedom reflects
the existence of the conformal Killing fields of $S^3$.

The lowest mode, $k = 1$,
corresponds to the negative eigenvalue
$\sigma_{(1)}^{2} = -2$ and gives rise to
exponentially growing perturbations.
As is seen from the above solution,
the metric perturbations depend
only on the time coordinate and the 
YM amplitude is not perturbed at all for 
the unstable mode,
\be
\da \, = \, \db \, = \, \dmu \, = \, e^{\sqrt{2} \, t} \, ,
\; \; \; \; \dw \, = \, 0 \, . 
\label{ground-8}
\ee

The second lowest mode, i.e. the mode with $\sigma^{2}_{(2)} = +1$,
is a pure gauge mode, since 
$\dzeta_{(k=2)} = \cos^{2} \! \rho 
(\frac{\sin(2\rho)}{\sin \rho 
\cos \rho})' = 0$. In fact, we find from eqs. (\ref{ground-4}), 
(\ref{ground-5}) and (\ref{ground-6}) for $k = 2$:
\bdm
\dmu = 2 \cos \rho \; e^{it} \, , \; \; \; 
\dw = - 2 \sin^{2} \! \rho \; e^{it} \, , \; \; \; 
\dc' = 0 \, .
\edm
A gauge transformation with $f = -2 \sin \rho \, e^{it}$ then gives
$\dmu = \dw = \dc' = 0$.

\subsection{An Alternative Approach}

The spectrum of perturbations of the ground state is also
obtained without using the gauge invariant
quantity $\dzeta$ as follows: Since the
spatial part of the background solution is a space
of constant curvature (the geometrical $3$--sphere),
it is natural to introduce the two gauge invariant 
Bardeen potentials used in cosmology to describe 
scalar perturbations \cite{JB}
(see, e.g., \cite{MFB} for a review). 
The Bardeen potentials are
proportional to $\da$ and $\db = \dmu$. We are
therefore again in the conformal gauge.
The fact that the norm of the
static Killing field of the background solution
is constant ($a = 0$) implies that the remaining
gauge freedom can be used to set $\dc = 0$, i.e.
$\da = \db$. (This is a peculiarity of
the ground state, for which
the equation for $\dc'$ decouples and
implies that $\dc'$ is actually a pure gauge; see above.) 
Hence, by virtue of the special form
of the background solution, there exists only
one independent scalar Bardeen potential. It is therefore
consistent to consider perturbations of the form
\be
\da \, = \db \, = \, \dmu \, ,
\; \; \; \; \; 
\dmu' \, = \, \dx \, ,
\label{C-1}
\ee
where the last equation is the constraint 
(\ref{constr-lin}) in the gauge
$\db = \dmu$ (with $a' = 0$,
$R = - w' = \sin \rho$.)

Due to the conformal invariance in four dimensions,
the metric perturbations do not enter the linearized
YM equation (\ref{YM-lin2}) for $\da = \db = \dmu$,
\be
\left[ \dal \, - \, 2 \, \frac{\cos^{2} \! \rho}{\sin^{2} \! \rho}
\right] \, \dx \, = \, 0 \, .
\label{C-2}
\ee
By virtue of the relations (\ref{C-1}), 
we obtain the following pulsation equation for $\dmu$ from
eq. (\ref{lin-1})
\be
\left[ \dal \, + \, 2 \, \right] \, \dmu \, = \, 0 \, .
\label{C-3}
\ee
Now using $(\dal \dmu)' = \dal \dmu' + 2 (\frac{\cos \rho}{\sin \rho})' 
\dmu' = (\dal -\frac{2}{\sin^{2} \! \rho}) \dmu'$ and $\dmu' = \dx$,
shows that the pulsation equation for $\dw$ is the derivative
of the pulsation equation for $\dmu$:
\bdm
\left( \, \left[ \dal \, + \, 2 \right] \, \dmu \,  \right)' \, = \, 
\left( \, \dal \, - \, 2 \, \frac{\cos^{2} \! \rho}{\sin^{2} \! \rho}
\, \right) \, \dx \, = \, 0 \, .
\edm

Since the background solution is a geometrical $3$--sphere, the 
$4$--dimensional d'Alembertian becomes
$\dal = \sin^{-2} \! \rho \, \partial_{\rho} (\sin^{2} \! \rho \,
\partial_{\rho})$. The solutions of eq. (\ref{C-3})
are therefore the harmonics of $S^{3}$,
\be
\dmu \, = \, \frac{\sin((\ell + 1) \rho)}{\sin \rho} \;
e^{i \sigma_{\ell} t} \, ,
\label{C-4}
\ee
where the spectrum is shifted 
by $-2$:
\be
\sigma^2_{\ell} \, = \, \ell \, (\ell+2) \, - \, 2 \, = \, 
(\ell + 1)^{2} \, - 3 \, .
\label{spec}
\ee
Since the unstable mode ($\ell = 0$) is purely time dependent,
$\dmu = e^{\sqrt{2} t}$, and since the
YM equation (\ref{C-2}) is the derivative of eq.
(\ref{C-3}), the unstable mode is missing in the spectrum of the
YM equation.
Using $\dw / \sin \rho = \dx = \dmu'$, $\mu' = \cos \rho / \sin \rho$
and $w' = -\sin \rho$, we finally find
\bea
\dzeta & = & \mu' \dw - w' \dmu \, = \,
\cos \rho \cdot \dmu' \, + \, \sin \rho \cdot \dmu
\nonumber \\
& = & \cos^{2} \! \rho \, \left( \frac{\dmu}{\cos^{2} \! \rho} \right)' 
\, = \, 
\cos^{2} \! \rho \, \left(\frac{\sin ((\ell + 1) \rho)}{\sin \rho 
\cos \rho} \right)' 
\, ,
\nonumber \\
\eea
which - with $k = \ell + 1$ - is 
in agreement with the results presented in
the previous section. 

\subsection{SUSY Transformation}

The potential appearing in the pulsation equation 
(\ref{YM-dw}) for $\dw$ (or in the 
corresponding equation (\ref{zeta-eq-2}) for $\dzeta$)
can diverge at the equator. 
(We have argued earlier that the equator is
a regular singular point.)
A pulsation equation with 
bounded potential can be obtained after a
supersymmetric transformation. This requires, however,
the knowledge of at least one solution of the
original equation. Unfortunately, we were not
able to find such a special solution in the general case. 
Nevertheless, the method
works perfectly for the ground state.

In order to demonstrate this, we consider the
pulsation equation (\ref{YM-dw}) for $\dw$. Since this
equation was derived in the Schwarzschild gauge, $\dmu = 0$,
and since 
\be
\dzetabar \, = \, \frac{1}{\mu'} \, \dzeta \, = \,
\dw \, - \, \frac{w'}{\mu'} \, \dmu
\label{susy1}
\ee
is clearly gauge invariant, eq. (\ref{YM-dw}) is also the correct
equation for $\dzetabar$. For the $S^3$ background solution
we have $\daltil = -\partial^{2}/\partial t^{2}+
\partial^{2}/\partial \rho^{2}$, and therefore
\be
\left( \drhotwo \, - \, {\cal P} \right) \, \dzetabar
\, = \, - \, \sigma^{2} \, \dzetabar \, ,
\label{susy2}
\ee
where the potential diverges at the equator, $\rho_{e} = \pi /2$,
\be
{\cal P} \, = \, \frac{3 \cos^{2} \! \rho -1 }{\sin^{2} \! \rho}
 \, + \, \frac{2}{\cos^{2} \! \rho} \, .
\label{susy2a}
\ee
Let $\psi$ be a particular solution of eq. (\ref{susy2}) with
eigenvalue $\sigma^{2}_{\psi}$, say. Then
the differential operators $\Bpl$ and $\Bmi$ are
defined as
\be
\Bpm \, = \, \drhoone \, \pm \, \frac{\psi'}{\psi} \, .
\label{susy3}
\ee
Now using ${\cal P} = \sigma^{2}_{\psi} + \psi''/ \psi$ and
$\Bpl \Bmi = \partial^{2}/\partial \rho^{2} - \psi''/ \psi$, 
eq. (\ref{susy2}) assumes the form
\be
\Bpl \, \Bmi \; \dzetabar \, = \, ( \, \sigma^{2}_{\psi} 
\, - \, \sigma^{2} \, ) \, \dzetabar \, .
\label{susy4}
\ee
Multiplying with $\Bmi$
gives
\be
\Bmi \, \Bpl \; \detapsi \, = \, ( \, \sigma^{2}_{\psi} 
\, - \, \sigma^{2} \, ) \, \detapsi \, ,
\label{susy5}
\ee
where $\detapsi$ is the 
supersymmetric partner of $\dzetabar$ with respect to
$\psi$,
\be
\detapsi \, = \, \Bmi \, \dzetabar \, = \, \dzetabar' \, - \, 
\frac{\psi'}{\psi} \, \dzetabar \, .
\label{susy6}
\ee
Since $\Bmi \Bpl = \partial^{2}/\partial \rho^{2} + (\psi' / \psi)'
- (\psi' / \psi)^{2}$, eq. (\ref{susy5}) finally becomes
\be
\left[ \drhotwo \, - \, {\cal P} \, + \, 
2 \left( \frac{\psi'}{\psi} \right)'
\right] \, \detapsi
\, = \, - \, \sigma^{2} \, \detapsi \, .
\label{susy7}
\ee
Hence, the potential in eq. (\ref{susy2}) 
picks up the additional contribution
$-2(\psi' / \psi)'$ which can compensate the 
divergent term in ${\cal P}$.
(It is obvious from the above derivation that
the eigenvalue $\sigma_{\psi}^{2}$ does not lie
in the spectrum of
the equation for the supersymmetric partner with respect
to $\psi$.)

As an example, we consider for $\psi$ the lowest mode,
$\psi = \dzetabar_{(1)} = \dzeta / \mu' = \sin^{2} \! \rho / \cos \rho$.
The effective potential for $\detapsi$ then becomes finite at the 
equator,
\be
{\cal P} \, - \, 2 \left( \frac{\psi'}{\psi} \right)' \, = \, 
3 \, \frac{ 1 + \cos^{2} \! \rho}{\sin^{2} \! \rho} \, .
\label{susy8}
\ee
Since we have performed a supersymmetric transformation
with respect to  $\dzetabar_{(1)}$, the unstable
mode is missing in eq. (\ref{susy7}) with potential (\ref{susy8}). 
In order to obtain this mode from an equation with finite potential, we
also consider the transformation with respect to the third
mode, $\dzeta_{(3)}$ (recall that $\dzetabar_{(2)} = 0$):
\bdm
\psi \, = \,  \frac{1}{\mu'} \dzeta_{(3)} \, = \,
\sin \rho \, \cos \rho 
\, \left( \frac{\sin (3 \rho)}{\sin \rho \, \cos \rho}
\right)' \,  = \, \frac{(c^2-1)(1 +4 c^2)}{c} \, , 
\edm
where $c \equiv \cos \rho$.
A straightforward calculation now yields 
the supersymmetric potential
\be
{\cal P} \, - \, 2 \left( \frac{\psi'}{\psi} \right)' \, = \, 
\frac{48 c^6 - 24 c^4 + 139 c^2 - 13}{(1-c^2)(1 + 4 c^2)^2} \, ,
\label{susy10}
\ee
which is bounded in the vicinity of the equator.
The pulsation equation (\ref{susy7}) with potential
(\ref{susy10}) has the unstable mode
\be
\detapsi \, = \frac{8 \, \sin^3 \! \rho}{1 + 4 \, \cos^2 \! \rho}
\label{susy11}
\ee
with negative eigenvalue $\sigma^2 = -2$. (It is now an
easy task to verify that $\detapsi$ is indeed the supersymmetric
partner of $\dzeta_{(1)}/\mu' = \sin^2 \! \rho / \cos \rho$
with respect to $\psi$,
\bdm
\detapsi = \left(\frac{s^2}{c}\right)' \, - \, 
\frac{\left( \frac{s^2}{c}(1+4 c^2) \right)'}
{\left( \frac{s^2}{c}(1+4 c^2) \right)} \, \frac{s^2}{c} \, ,
\edm
where $s = \sin \rho$.)
Hence, we have obtained the unstable mode from
a perfectly regular pulsation equation in standard form.
%
%

\section{The Odd-Parity Sector}

In order to study purely magnetic
solutions it is sufficient to
work with a gauge potential of the form
(\ref{YM-1form}). However, if one is interested
in the stability properties of such configurations,
one must, in general, consider perturbations
of the full spherically symmetric gauge potential
\be
A \, = \, \tilde{A} \, \hat{\tau}_{\rho} 
\, + \, \varpi \, 
(\hat{\tau}_{\vartheta} d \vartheta + 
\hat{\tau}_{\varphi} \sin \vartheta d \varphi)
\, + \, (w-1) \,
(\hat{\tau}_{\varphi} d \vartheta - 
\hat{\tau}_{\vartheta} \sin \vartheta d \varphi) \, ,
\label{o2}
\ee
where $\tilde{A} = a_0 dt + a_1 d \rho$.
It is a crucial observation that 
$\dAtil$ and $\dpi$ decouple from the remaining perturbations, that is,
the equations for $\dAtil$ and $\dpi$ 
contain no metric perturbations and no $\dw$'s -- and
vice versa. This is a
consequence of the fact that the parity transformation 
\be
\vartheta \, \rightarrow \, \pi \, - \, \vartheta \, ,
\ \ \ \ \ \ 
\varphi \, \rightarrow \, \pi \, + \, \varphi 
\label{o1}
\ee
acts as a symmetry operation on the background solutions.
Until now, we have restricted ourselves to the fluctuations
with even parity, that is, we have considered perturbations
for which $\dAtil = 0$ and $\dpi = 0$.

In this section we shall investigate the orthogonal set of
fluctuations that is, we discuss the pulsation equations
for $\dAtil$ and $\dpi$ on a static, purely magnetic
background.
We do so by taking advantage of a method developed in 
\cite{VBLS}. 
The following analysis
applies to all solutions of the EYM equations 
with cosmological constant studied in \cite{VSLHB}. 
The reason for this lies in the fact that it is actually 
sufficient to consider perturbations
which vanish at the horizon (see \cite{VG}
for details.)
 
\subsection{Pulsation Equations}

The YM equations for the general spherically symmetric
gauge potential (\ref{o2}) are most easily obtained from the YM action, 
$S = \int_M \mbox{tr} (F \we \ast F)$. Using the general form
$\gtens = \gtiltens + R^2 \ghattens$ of the metric and the
fact that $w$, $\varpi$ and $R$ are functions on $\tilde{M}$,
whereas $\tilde{A}$ is a $1$--form on $\tilde{M}$, 
one finds
\bdm
S = \int_{\tilde{M}} \tilde{\eta}
\left[
\frac{R^2}{2} \mid d \tilde{A} \mid ^2 +
\mid d \varpi - \tilde{A} w \mid ^2 +
\mid d w + \tilde{A} \varpi \mid ^2 +
\frac{V(w,\varpi)}{2 R^2}
\right] \, ,
\edm
where $V(w,\varpi)=(\varpi^2 + w^2 -1)^2$ and
$\mid d \tilde{A} \mid ^2 \tilde{\eta} =
d \tilde{A} \we \tilde{\ast} d \tilde{A}$,
$\tilde{\eta}= \sqrt{\tilde{g}} \, dt \we dr$.
Now using
\bdm
\mid dt \mid^2 = -e^{-2a} \, ,
\; \; \; \; 
\mid dr \mid^2 = e^{-2b} \, ,
\; \; \; \; \mbox{and} \; \; 
\tilde{A} = a_0 dt + a_1 d\rho \, ,
\edm
we obtain the YM equations upon performing variations of 
$S$ with respect to $a_0$, $a_1$, $\varpi$ and $w$.
The linearization of these equations on a static background
with $a_0=a_1=\varpi=0$ finally gives
\bea
\left[R^2 e^{-(a+b)} (\daoned - \dazer') \right]' & = &
-2 e^{b-a} \left[ w^2 \dazer - w \dpidot \right] \, ,
\label{odd-lin-1}\\
\left[R^2 e^{-(a+b)} (\daoned - \dazer') \right] \dot{} & = &
-2 e^{a-b} \left[ w^2 \daone - w \dpi' + w' \dpi \right] \, ,
\label{odd-lin-2}
\eea
\bdm
- \, \left[ e^{b-a} (\dpidot - w \dazer) \right] \dot{} \, + \, 
\left[ e^{a-b} (\dpi' - w \daone) \right]'
\edm
\be
= \, e^{a-b} w' \daone \, + \, e^{a+b} \, \frac{w^2-1}{R^2} \, \dpi \, .
\label{odd-lin-3}
\ee
In addition, we have the linearized YM equation (\ref{YM-lin1})
for $\dw$, which remains unchanged.
As we already pointed out, the above three equations for the
fluctuations with odd parity do neither involve the metric
perturbations nor the even parity quantity $\dw$.

We now adopt the temporal gauge and
consider the following separation ansatz for the perturbations:
\be
\dazer = 0, \ \ \
\daone = \frac{2e^{a+b}}{R^2} \, \dchi(\rho) \, e^{i \sigma t},\ \ \
\dpi = \dxi(\rho) \, e^{i \sigma t}.\ \ \
\label{o4}
\ee
Also introducing the background quantity $\gamma$ and the new coordinate
$x$,
\be
\gamma^2 \, = \, \frac{2}{R^2} \, e^{2a} \, , \ \ \ \ 
dx \, = \, e^{b-a} \, d \rho \, ,
\label{08}
\ee
yields the following set of ordinary
differential equations for $\dchi$ and $\dxi$:
\be
\sigma \, ( \dchi,_x \, - \, w \, \dxi) \, = \, 0 \, ,
\label{o12}
\ee
\be
w^{2}\gamma^{2} \dchi \, + \, w,_x \, \dxi \, - \, 
w \, \dxi,_x \, = \, \sigma^2 \, \dchi \, ,
\label{o11}
\ee
\be
\left(
w^{2}\gamma^{2} \dchi \, + \, w,_x \, \dxi \, - \, 
w \, \dxi,_x \right),_x \, = \, \sigma^2 \, w \, \dxi \, .
\label{o10}
\ee
In the last equation we have also used the 
background YM equation for $w$ in the form 
$2w,_{xx} = \gamma^2 w (w^2-1)$.
Equation (\ref{o12}) is an obvious consequence
of the first two equations. In order to gain a 
second order equation from the first order
equations (\ref{o12}) and (\ref{o11}), we
introduce the function $\dphi \equiv \dchi / w$,
for which we find the Schr\"odinger equation
\be
- \, \dphi,_{xx} \, + \, \left[
 \frac{1}{2} \gamma^2 (w^2-1) + 2 \left(
\frac{w'}{w} \right)^2 \right] \, \dphi \, = \, \sigma^2 \, \dphi \, .
\label{schr-phi1}
\ee
(For $\sigma^2 \neq 0$ this equation is equivalent to
the system (\ref{o12})--(\ref{o10}). For $\sigma^2 = 0$
we impose eq. (\ref{schr-phi1}) as a gauge fixing condition; 
see \cite{VBLS}.)
A particular solution -- which is related to the
residual gauge freedom for $\sigma^2 = 0$ -- is (see \cite{VSLHB})
\be
\dphi_0 \, = \, \frac{\Omega,_x}{\gamma^2 \, w} \, ,
\label{schr-phi2}
\ee
where the function $\Omega(x)$ is subject to the
differential equation
\be
\left( \frac{1}{\gamma^2} \, \Omega,_x \right),_x \, = \, w^2 \, \Omega \, .
\label{o19}
\ee

\subsection{The Number of Odd-Parity Instabilities}

The potential in the Schr\"odinger equation (\ref{schr-phi1})
is non--negative and becomes unbounded at the zeroes of the
background amplitude $w$. This suggests that the number of nodes
of $w$ determines the number of eigenfunctions with negative
$\sigma^2$. In fact, in \cite{VBLS} we have shown that this
is the case for the BK solitons and the corresponding
black hole solutions. We shall now take advantage of
the technique introduced in \cite{VBLS} to show that the number
of nodes of $w$ equals the number of unstable modes
for all solutions with cosmological constant
described in \cite{VSLHB}.

The key idea is to apply a 
supersymmetric transformation of eq. (\ref{schr-phi1}) 
with respect to the zero--energy solution $\dphi_0$ given in
eq. (\ref{schr-phi2}). The supersymmetric partner
$\dpsi$ of $\dphi$ then fulfills 
a similar Schr\"odinger equatian as $\dphi$, where
the additional term $-2 \dphi_0,_x/\dphi_0$ is added
to the potential in eq. (\ref{schr-phi2}) 
(see also sec. 5.3, eqs. (\ref{susy2}), (\ref{susy7})).
In \cite{VBLS} we have demonstrated that 
the pulsation equation for $\dpsi$ can then be written in 
the form
\be    
- \, \dpsi,_{xx} \, + \, \left[
\frac{1}{2}\gamma^{2}
(3w^{2}-1)+2(w^{2}Z),_x \right] \, \dpsi \, = \, \sigma^{2} \,  \dpsi,
\label{o15} 
\ee
where $Z$ is obtained from
\be
Z(x) \, = \, - \frac{\gamma^2}{\bar{\Omega},_x} \, \left[
\bar{\Omega} \, + \,  \frac{\gamma^2 / \bar{\Omega},_x}{C+
{\displaystyle\int_{0}^{x}}(w \gamma^2 / \bar{\Omega},_x)^2 \, d \bar{x}} 
\right] \, .
\label{o18}
\ee
Here $C$ is an arbitrary constant and
$\bar{\Omega}$ denotes a {\em particular\/} solution
of the differential equation (\ref{o19}).
For $\sigma=0$, eq. (\ref{o15}) admits the solution
\be
\dpsi_{0}(x) \, = \, w \, \exp\left(\int_{x_{0}}^{x}w^{2}Z \,
d \bar{x}\right) \, .
\label{o17}
\ee
If $Z$ is everywhere regular, then $\dpsi_{0}$ is also smooth and
regular. This implies that $\dpsi_0$ has $n$ zeroes, which will enable us
to count the number of bound states. Hence, we have to
investigate the behavior of $\bar{\Omega}$, from which
we obtain $Z$ and $\dpsi_{0}$ by virtue of eqs.
(\ref{o18}) and (\ref{o17}), respectively.

In order to analyze the above equations, it is convenient 
to choose the background gauge
\be
e^{2a(\rho)} \, = \, e^{-2b(\rho)} \, \equiv \, Q(\rho) \, ,
\label{o20}
\ee
which we have also used in \cite{VSLHB}.
Also returning to the radial coordinate $\rho$
($dx=d\rho/Q$, $\gamma^2=2Q/R^2$), the differential equation
(\ref{o19}) for $\Omega$ becomes 
\be
Q \, \left( R^2 \, \Omega' \right)' \, = \, 
2 \, w^{2} \, \Omega \, .
\label{o21}
\ee

Our first aim is to show that -- for all
static background solutions with horizon  
discussed in \cite{VSLHB} -- there exists a
smooth solution $\bar{\Omega}(\rho)$ 
of eq. (\ref{o21}) which behaves as
\be
\bar{\Omega}(\rho) \, = \, \frac{k}{\rho^2} \, + \, 
{\cal O}(1) \, , \; \; \; \mbox{and} \; \; 
\bar{\Omega}(\rho) \, = \, {\cal O}(\rho - \rho_h) \, , 
\label{o28}
\ee
in the vicinity of the origin ($\rho = 0$) and the
horizon ($\rho = \rho_h$), respectively.
Moreover, $\bar{\Omega}^2$
is monotonically decreasing between $\rho = 0$
and $\rho = \rho_h$.

In order to discuss the behavior near the origin,
we recall that all solutions 
obtained in \cite{VSLHB} behave as
(see \cite{VSLHB}, eq. (49))
\be
w \, = \, 1 + {\cal O}(\rho^{2}),\ \ \ \
R \, = \, \rho + {\cal O}(\rho^{5}),\ \ \ \ 
Q \, = \, 1 + {\cal O}(\rho^{2}),                           
\label{o22}
\ee
for $\rho \rightarrow 0$.
Hence, the origin is a regular singular point
of eq. (\ref{o21}) with roots $r_{\pm} = 1,-2$.
In the vicinity of the horizon we have
(see \cite{VSLHB}, eq. (51))
\be
w \, = \, w_h + {\cal O}(\delrho),\ \ \ \
R \, = \, R_h + {\cal O}(\delrho), \ \ \ \ 
Q \, = \, |Q'_h| \, \delrho + {\cal O}(\delrho^2),\ \ \ \
\label{o23}
\ee
where $\delrho \equiv \rho_h - \rho$ and
$Q'_h < 0$. 
Using this in eq. (\ref{o21})
shows that there exists a local solution 
with $\bar{\Omega}(\rho) = {\cal O}(\delrho)$.
Finally, in order to establish
the monotonicity property, we multiply
eq. (\ref{o12}) with $\Omega / Q$ and
integrate from $\rho < \rho_h$
to $\rho$. Using $\bar{\Omega}(\rho_h) = 0$,
this yields the desired result:
\be
\left(\bar{\Omega}^{2}\right)' (\rho) \, = \, 
-\frac{2}{R^{2}(\rho)} \, \int_{\rho}^{\rho_{h}}
\left(R^{2} \, \bar{\Omega}'^{2} \, + \, 
2\frac{w^{2}}{Q} \, \bar{\Omega}^{2}\right) d\bar{\rho} \, < \, 0 \, .
\label{o27}
\ee

Using the expansions for $Q$ and the definition
$dx = d \rho / Q$ of the coordinate $x$, we find 
$x = \rho + {\cal O}(\rho^3)$ as $\rho \rightarrow 0$
and $x = -|Q'_h|^{-1} \ln (\delrho) + {\cal O}(1)$ 
as $\rho \rightarrow \rho_h$. Hence, eq. (\ref{o28})
implies
\be
\bar{\Omega}(x) \, = \, \frac{k}{x^2} \, + \, 
{\cal O}(1) \, , \; \; \; \mbox{and} \; \; 
\bar{\Omega}(x) \, = \, 
{\cal O}(\exp(-|Q'_h| x))
\label{o28bis}
\ee
in the vicinity of the origin ($x = 0$) and
the horizon ($x = \infty$), respectively.
Using this in the expression (\ref{o18}) for the 
function $Z(x)$ shows that
\be
Z(x) \, = \, \frac{1}{x} + {\cal O}(x)
\, , \; \; \; \mbox{and} \; \; 
Z(x) \, = \, - \, w^2_h \, \frac{1}{x} + {\cal O}(\frac{1}{x^2})
\label{o29}
\ee
as $x \rightarrow 0$ and $x \rightarrow \infty$, respectively.
Finally, by virtue of eq. (\ref{o17}), we find for the
solution $\dpsi_0(x)$ of eq. (\ref{o15})
\be
\dpsi_0(x) \, = \, c_0 \, x \, + \, {\cal O}(x^2)
\, , \; \; \; \mbox{and} \; \; 
\dpsi_0(x) \, = \, c_{\infty} \, \frac{1}{x} 
\, + \, {\cal O}(\frac{1}{x^2})
\label{o30}
\ee
as $x \rightarrow 0$ and $x \rightarrow \infty$, respectively.
This shows that $\psi_{0}\in L^{2}(0,\infty)$. 
Since $\psi_{0}$
is the zero--energy solution of eq. (\ref{o15}), and since
$\psi_{0}$ has the same number of nodes as the background solution $w$
on the interval $(0,\infty)$,
we conclude that eq. (\ref{o15}) has $n$ bound states,
$\psi_{k}(x)\in L^{2}(0,\infty)$
with $\sigma_{k}^{2}<0$, $k=1\ldots n$.

For $\sigma\neq 0$, the supersymmetric relation
between $\dphi$ and $\dpsi$ (with respect to $\dphi_0$)
guarantees that eq. (\ref{schr-phi1}) has also
$n$ bound states. Finally since the Schr\"odinger
equation for $\dphi$ is equivalent to the system
(\ref{o12})--(\ref{o10}),
the same holds for the solutions
$\dchi$ and $\dxi$. Explicitly, one finds
(see \cite{VBLS})
\be
\dchi_k \, = \, \frac{1}{\sigma^{2}}\left(w^{3}Z\dpsi_k+
w' \dpsi_k - w \dpsi'_k \right) ,\ \ \ \ \ \ \
\dxi_k \,  = \, \dpsi_k - w Z \, \dchi_k \, . 
\label{o31}
\ee
This proves the existence of exactly $n$ unstable modes 
in the spherically--symmetric, odd--parity sector for 
all EYM solutions with 
$0<\Lambda <\Lambda_{\star}(n)$ and
$\Lambda_{\star}(n)< \Lambda <\Lambda_{\mbox{\scriptsize{reg}}}(n)$ 
obtained in \cite{VSLHB}.
(The special case $\Lambda = \Lambda_{\star}(n)$,
for which the horizon occurs at the maximum $\rho_e$ of $R$,
is not covered by the above reasoning, since
eq. (\ref{o21}) develops an {\it irregular} singular
point at $\rho = \rho_e = \rho_{h}$. Although we have not carried out
the analysis for this case, we expect on continuity grounds
that the result remains the same.)
In the limit $\Lambda\rightarrow\Lambda_{\mbox \scriptsize{reg}}(n)$, 
corresponding to the compact solutions, 
the horizon shrinks to a point (the ``south pole''), 
where the boundary conditions are 
identical with the boundary conditions at the origin.
The above analysis
can easily be repeated for this situation, and reveals 
again $n$ unstable modes for
the compact solutions with $n$ zeroes.

\section{Concluding Remarks}

In this paper we have investigated the stability of the static,
purely magnetic EYM solutions with regular center 
and positive cosmological constant. Whilst the numerical analysis is 
straightforward for the asymptotically deSitter solutions, new
difficulties arise for the compact and the bag of gold type
configurations. 

In order to solve these problems, we have
generalized an elimination procedure
for the gravitational fluctuations -- earlier
derived for the Schwarzschild gauge ($\dR = 0$) --
in a gauge invariant manner. In this way, we end up with 
a standard Sturm--Liouville equation for a gauge invariant
quantity $\dzeta$. 
We have finally argued that there exists a globally
regular gauge, with respect to which the metric and matter 
fluctuations can be reconstructed from $\dzeta$.

The numerical investigation 
for the first few branches ($n=1,2,3$)
shows that there are $n$ unstable modes
for every solution with $n$ zeroes of $w$
(in the even parity sector).
This turns out to be true for the
entire interval $[0,\Lregn]$, that is, for
the BK solutions, their asymptotically
deSitter--like analogues, the bag of gold
solutions and the compact configurations.

Adopting a method originally developed for the BK soliton and black
hole solutions, we have finally also established 
analytically the existence of $n$ 
unstable modes in the odd parity sector. The procedure, 
which is based on 
a supersymmetric transformation of the Schr\"odinger equation
for the sphaleron--like modes, was explicitly carried out for the bag of
gold solutions. As expected, the compact solutions also exhibit
$n$ unstable modes with odd parity.
Hence, we conclude that every solution -- characterized by the
number of zeroes $n$ and the cosmological constant
$\Lambda(n) \in [0,\Lregn]$ -- has exactly $2n$ unstable modes.

\end{document}